\documentclass[twocolumn, prx, reprint, amsfonts, amsmath, amssymb, longbibliography, floatfix, aps]{revtex4-2}

\usepackage[T1]{fontenc}
\usepackage{lmodern}
\usepackage{graphicx}
\usepackage[dvipsnames]{xcolor}
\usepackage[colorlinks,linkcolor=Blue,citecolor=Blue]{hyperref}
\usepackage{braket}
\usepackage{mathtools}
\usepackage{enumitem}
\usepackage{microtype}

\usepackage{booktabs}
\usepackage{array}
\usepackage{makecell}
\usepackage[normalem]{ulem}

\usepackage{cancel}

\begin{document}

\title{Dissipation-Shaped Quantum Geometry in Nonlinear Transport}

\author{Zhichao Guo}
\author{Xing-Yuan Liu}
\author{Hua Wang}
\author{Li-kun Shi}
\email{likun.shi@zju.edu.cn}
\author{Kai Chang}
\email{kchang@zju.edu.cn}

\affiliation{Center for Quantum Matter, School of Physics, Zhejiang University, Hangzhou 310058, China}

\date{\today}

\begin{abstract}
The theory of the intrinsic nonlinear Hall effect, a key probe of quantum geometry, is plagued by conflicting expressions for the conductivity that is independent of the dissipation {\it strength} (rate, $\Gamma^0$).
We clarify the origin of this ambiguity by demonstrating that the ``intrinsic'' response is not universal, but is inextricably linked to how the dissipation {\it mechanism} shapes the non-equilibrium steady state (NESS) density matrix.
We establish a benchmark by solving the exact NESS density matrix for a generic Bloch system coupled to a featureless fermionic bath. 
Our exact $\Gamma^0$ conductivity decomposes into two parts: (i) a geometric contribution, $\sigma^{\text{geo}}$, which establishes the definitive structure of the quantum metric contribution (including the intraband $\sim \partial_k g$ term), clarifying inconsistencies in the literature, and (ii) a novel, purely kinetic contribution, $\sigma^{\text{kin}} \propto v^3 f^{(4)}_0$, arising from mechanism-specific modifications to the occupation functions, which is absent in approaches that postulate, rather than derive, the relaxation dynamics.
The discrepancies in both $\sigma^{\text {geo }}$ and $\sigma^{\text {kin }}$ between these distinct physical mechanisms prove that the $\Gamma^0$ nonlinear conductivity is not a unique property of the Bloch Hamiltonian, but is contingent on the physical system-bath coupling.
\end{abstract}

\maketitle

\bigskip

\textit{\color{blue}Introduction.}
Linear response theory establishes a profound connection between transport phenomena and the properties of the thermodynamic equilibrium state, which are insensitive to the specific nature of the environment, as exemplified by the relationship between the intrinsic anomalous Hall effect and the Berry curvature~\cite{thouless1982quantized,berry1984quantal,haldane1988model,sinitsyn2007semiclassical, nagaosa2010anomalous,xiao2010berry,resta2011insulating}. 
This paradigm, however, breaks down in the nonlinear regime.
DC nonlinear responses are governed not by equilibrium properties, but by the non-equilibrium steady state (NESS). 
The structure of the NESS, i.e., the non-equilibrium density matrix, emerges from a delicate balance between the external drive and dissipation.

Importantly, we must distinguish between the {\it mechanism} of dissipation (the microscopic nature of the system-environment coupling, e.g., metallic backgate, phonons, or disorder statistics) and the {\it strength} of dissipation (the phenomenological rate, $\Gamma$). 
The mechanism determines the analytical structure of the NESS density matrix (e.g., how occupation functions are modified) \cite{tien1963multiphoton, mani2002zero,zudov2003evidence,bukov2015universal,sentef2017light,oka2019floquet,kumari2024josephson,le2024inverse,eckhardt2024theory,matsyshyn2023fermi,zhu2024anomalous,shi2024floquet,shi2025ultracritical}, and is therefore central, not peripheral, to the resulting nonlinear transport.

Recent years have witnessed intense theoretical efforts to interpret nonlinear transport through the lens of quantum geometry~\cite{ QGT_Provost_1980,QGT_PATI_1991,gao2014field,sodemann2015quantum,peotta2015superfluidity,nagaosa2017concept,wang2019ferroelectric,matsyshyn2019nonlinear,ma2019observation,xie2020topology,watanabe2020nonlinear,watanabe2021chiral,wang2021intrinsic,liu2021intrinsic,cayssol2021topological,du2021quantum,oiwa2022systematic,torma2023essay,das2023intrinsic,kirikoshi2023microscopic,shi2023berry,wang2023quantum, gao2023quantum,hetenyi2023fluctuations, gao2024antiferromagnetic,wang2024intrinsic, kaplan2024unification, mehraeen2024quantum, liu2025quantum,zhu2025magnetic,qiang2025clarification,ulrich2025quantum, xiao2025proper,resta2025intrinsic, fontana2025quantum, mehraeen2025_gravity}. 
Many of these efforts aim to identify ``intrinsic'' contributions, which are independent of dissipation strength ($\Gamma^0$) and are characterized by the quantum metric~\cite{QGT_Provost_1980,thouless1982quantized,berry1984quantal,QGT_PATI_1991,peotta2015superfluidity, xie2020topology,resta2011insulating,cayssol2021topological,torma2023essay, xiao2025proper, resta2025intrinsic}. 
Despite significant progress, the field is plagued by inconsistencies. 
Different theoretical frameworks, including semiclassical wave-packet dynamics~\cite{gao2014field,wang2021intrinsic,liu2021intrinsic,qiang2025clarification}, quantum kinetics or Green's function methods~\cite{nagaosa2017concept,wang2019ferroelectric,matsyshyn2019nonlinear,watanabe2020nonlinear,watanabe2021chiral,du2021quantum,oiwa2022systematic,das2023intrinsic,kirikoshi2023microscopic,kaplan2024unification,wang2024intrinsic,zhu2025magnetic,ulrich2025quantum}, yield conflicting expressions for these ``intrinsic'' conductivities.
\begin{figure}
\includegraphics[width=0.48\textwidth]{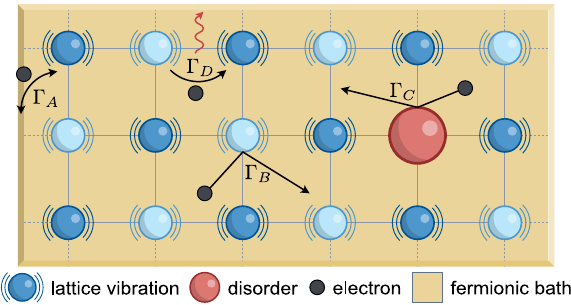}
\caption{Non-universality of the $\Gamma^0$ nonlinear conductivity. 
The non-equilibrium density matrix governing DC transport depends on the specific dissipation {\it mechanism}, leading to mechanism-dependent ``intrinsic'' $(\Gamma^0)$ conductivities. 
Illustrated environments mediating relaxation: $(\Gamma_A)$ fermionic bath (e.g., metallic backgate); $(\Gamma_B)$ electron-phonon interactions; $(\Gamma_C)$ static disorder; $(\Gamma_D$) radiative processes.}
\label{fig-schematic}
\end{figure}
\begin{table*}[t]
    \centering
    \footnotesize
    \begin{tabular}{ c | c | c | c }
        \toprule
        \toprule
        \textbf{Methods} & 
        \textbf{Dissipation Mechanism}& \textbf{$\Gamma^0$ Geometric} $\sigma^\text{geo}_{abc}$ & \textbf{$\Gamma^0$ Kinetic} $\sigma^\text{kin}_{abc}$ \\
        \midrule
        \makecell*[c]{Wavepacket or Liouville dynamics
        \\
        with RTAs or IFRs 
        \\
        (e.g., Refs.~\cite{gao2014field,das2023intrinsic,kaplan2024unification, qiang2025clarification})} &
        \makecell*[c]{Phenomenological
        \\
        $c_1/c_2$ varies depending
        \\on different assumptions}
        & \makecell*[c]{$\displaystyle 
        \sum_{n,{\bf k}}\left[ c_1 (G_n^{ab} v_n^c + G_n^{ac} v_n^b) - c_2 G_n^{bc} v_n^a \right] f_{0,n}'$} & \makecell*[c]{$0$}\\
        [0.4cm]
        \makecell*[c]{Green's function~\cite{ulrich2025quantum}} &
        \makecell*[c]{Constant-$\Gamma$
        \\self-energy}
        &   \makecell*[c]{$\displaystyle 
        \sum_{n,{\bf k}} \left[ (G_n^{ab} v_n^c + G_n^{ac} v_n^b - 2 G_n^{bc} v_n^a) -\frac{\partial_a g_n^{bc}}{2}  \right] f_{0,n}'$}  & \makecell*[c]{$0$}
        \\
        [0.4cm]
        \makecell*[c]{Open-system
        Schrödinger\\
        equation (this work)} &
        \makecell*[c]{Microscopic:\\fermionic bath}
        &  \makecell*[c]{$\displaystyle 
        \sum_{n,{\bf k}} \left[ (G_n^{ab} v_n^c + G_n^{ac} v_n^b - 2 G_n^{bc} v_n^a) -\frac{\partial_a g_n^{bc}}{2}  \right] 2 f_{0,n}'$}  &  \makecell*[c]{$ \displaystyle 
        \sum_{n,{\bf k}} (v^{a}_{n}v^{b}_{n}v^{c}_{n}) \frac{f^{(4)}_{0, n}}{24}$}\\
        [0.4cm]
        \bottomrule
        \bottomrule
    \end{tabular}
    \setlength{\abovecaptionskip}{0.4cm}
    \caption{Comparison of the ``intrinsic'' ($\Gamma^0$) contributions to the second-order DC conductivity $\sigma^{(0)}_{abc} = \sigma^\text{geo}_{abc} + \sigma^\text{kin}_{abc}$ derived from different theoretical approaches and the dissipation mechanisms they implicitly or explicitly model.
    The discrepancies in both $\sigma_{a b c}^{\text {geo }}$ (varying coefficients $c_1 / c_2$ and the intraband term $\partial_a g_n^{b c}$) and $\sigma_{a b c}^{\text {kin }}$ between the distinct dissipation mechanisms demonstrate the non-universality of the $\Gamma^0$ response. 
    Here, $g_n^{ab}$ ($G_n^{ab}$) is the (energy normalized) quantum metric for band $n$, $v_n^a$ is the band velocity, $f'_{0,n}$ and $f^{(4)}_{0,n}$ are the first and fourth derivatives of the equilibrium Fermi-Dirac distribution $f_0$. 
    Different conventions for symmetrizing $\sigma_{a b c}$ (or electron charge) can lead to an overall factor of 2 (or $-1$) difference.}
    \label{tab:comparison}
\end{table*}

We argue that these inconsistencies expose a fundamental physical misconception. 
The implicit assumption that the $\Gamma^0$ conductivity is universal, defined solely by the system Hamiltonian, is flawed. 
The NESS, which governs DC transport, is inextricably linked to the specific physical \textit{mechanism} of dissipation (Fig.~\ref{fig-schematic}). 
The analytical structure of the non-equilibrium density matrix, imposed by the mechanism dictates the precise procedure for taking the DC limit ($\omega \to 0$) followed by the $\Gamma$ expansion.

The reliance on phenomenological treatments in the literature obscures this fundamental dependence. 
Methods such as relaxation time approximations (RTAs) and imaginary frequency regularizations (IFRs) do not derive the analytical structure of the non-equilibrium density matrix
from microscopic principles; instead, they impose it via \textit{ad hoc} assumptions (e.g., regarding how occupations relax). 
The resulting $\Gamma^0$ conductivity therefore depends on these choices. For instance, RTA implementations differ on the postulated physical target state for relaxation: some assume relaxation towards the equilibrium distribution of the unperturbed Hamiltonian, $f_0(\epsilon)$~\cite{kaplan2024unification}, while others assume relaxation towards a quasi-equilibrium state defined by the field-modified energy (e.g., incorporating the Stark shift), $f_0(\tilde{\epsilon})$~\cite{gao2014field, qiang2025clarification}.

Further ambiguity arises in the regularization procedure required for the DC limit, e.g., there is no consensus on whether the regularization factor should remain constant, $i\Gamma$~\cite{watanabe2020nonlinear,watanabe2021chiral}, or depend on the order of the perturbation, $iN\Gamma$~\cite{das2023intrinsic,kaplan2024unification,wang2024intrinsic,zhu2025magnetic}, sometimes rationalized as a mathematical device for adiabatic switching. 
These technical differences change how divergences are regularized when taking the $\omega \to 0$ followed by $\Gamma$ expansions. 
This proliferation of approaches, based on distinct phenomenological assumptions, leads to the conflicting expressions summarized in Table~\ref{tab:comparison} (varying $c_1/c_2$), underscoring the inadequacy of theories that do not specify the physical dissipation mechanism.

In this Letter, we provide a theoretical benchmark by analyzing a concrete, microscopically defined open quantum system: a crystalline electronic system coupled to a wide-band fermionic bath. 
While idealized, this model allows for a controlled derivation of the NESS, capturing the essential physics of particle exchange with a Markovian fermionic environment (e.g., metallic backgate)~\cite{seetharam2015controlled,morimoto2016topological,matsyshyn2021rabi,shi2024floquet, shi2025ultracritical,shi2023berry,matsyshyn2023fermi} and providing a necessary contrast to idealized phenomenological models.

Our analysis yields the definitive second-order conductivity $\sigma_{abc}$ for this specific mechanism.
We find that the $\Gamma^0$ contribution contains not only a specific form of the quantum metric contribution, comprising both interband and intraband (derivative) terms, but also a novel, $\Gamma^0$ kinetic term proportional to the fourth derivative of the Fermi function, $f_0^{(4)}$.
The specific analytical form of the geometric term and the appearance of the kinetic term demonstrate that the structure of the $\Gamma^0$ nonlinear conductivity is non-universal and depends on the nature of the system-bath coupling.

\medskip

\textit{\color{blue}Microscopic Open-System Formalism.} To avoid the ambiguities inherent in phenomenological treatments of dissipation, we analyze a concrete open quantum system where the regularization of the NESS is microscopically determined, rather than postulated. 
The total Hamiltonian we consider is $H(t) = H_S(t) + H_B + H_{SB}$. The system, $H_S(t)$, describes Bloch electrons driven by a spatially uniform electric field ${\bf E}(t)$. We employ the velocity gauge, incorporating the drive via the Peierls substitution (setting $e=\hbar=1$)
$H_S(t) = H_0({\bf k} - {\bf A}(t))$,
where $H_0 = h_{m n}({\bf k}) |\chi_m\rangle \langle\chi_n| $ is the unperturbed Bloch Hamiltonian with eigenstates $|\chi_n\rangle$ (incorporating both momentum and band indices) and ${\bf E}(t) = -\partial_t {\bf A}(t)$.
The environment is modeled as a fermionic bath~\cite{seetharam2015controlled, matsyshyn2023fermi, shi2023berry, matsyshyn2021rabi}. 
We adopt the bath Hamiltonian $H_B = \sum_{n,j}\epsilon_{j}|\varphi_{n,j}\rangle\langle\varphi_{n,j}|$ and the system-bath tunnel coupling $H_{SB} = \lambda\sum_{n,j}(|\chi_{n}\rangle\langle\varphi_{n,j}| + \text{h.c.})$.
This Bloch eigenstate coupling structure describes a translationally invariant interaction, enforced by momentum conservation when the system couples uniformly to the bath.
This model is physically motivated by van der Waals heterostructures; for instance, when a 2D electronic system couples uniformly to a metallic backgate (e.g., graphite) via an ultrathin tunnel barrier (e.g., $\sim 2$ nm hBN).
This model provides an elementary theoretical benchmark, distinct from spatially localized coupling (e.g., contacts at system edges), and enables a direct comparison with translationally invariant phenomenological models and a clear formulation of transport theory in terms of momentum-space quantum geometry.

The bath is initialized in thermal equilibrium characterized by the Fermi-Dirac distribution $f_0(\epsilon) = [e^{\beta(\epsilon-\mu)}+1]^{-1}$ with $1/\beta$ the bath temperature and $\mu$ the bath chemical potential. 
We employ the wide-band approximation, treating the bath's density of states as featureless, i.e., $\nu_B(\omega)=\nu_0$.
This model is in the same class of those non-interacting fermionic models often described within the Keldysh formalism~\cite{gerchikov1989theory,jauho1994time,johnsen1999quasienergy,kohler2005driven,kamenev2011field,nagaosa2017concept,matsyshyn2021rabi}, and captures the essential physics of a Markovian environment and yields a well-defined, energy-independent relaxation rate $\Gamma = \lambda^2 \nu_0/2$.

The dynamics of the system's reduced density matrix, $\rho_S(t)$, are determined by tracing out the bath degrees of freedom~\cite{shi2023berry,shi2024floquet}. 
To extract the nonlinear conductivity, we expand the system's reduced density matrix perturbatively with respect to the driving $\rho_S(t) = \rho^{(0)} + \rho^{(1)}(t) + \rho^{(2)}(t) + {\cal O}(V^3)$, where $V(t) = H_S(t) - H_0$.
The microscopic derivation yields exact expressions for the response kernels (detailed in the Supplemental Material~\cite{supplemental}). 
Their analytical structure differ significantly from those obtained via RTAs or IFRs.

Crucially, this specific structure determines how the fermionic bath regularizes divergences when taking the DC ($\omega \to 0$) limit followed by $\Gamma$ expansion. 
In RTAs, DC corrections to the distribution function typically diverge as powers of $1/\Gamma$. 
In contrast, the fermionic bath regularization leads to \textit{finite}, $\Gamma^0$ corrections to the NESS occupation functions even in the clean limit.
This distinction highlights a fundamental principle: the precise analytical form of these $\Gamma$-independent corrections is inherently determined by the physical dissipation mechanism. 
It is this mechanism-specific structure of the NESS in the clean limit that constitutes the origin of the non-universal $\Gamma^0$ conductivity~\cite{shi2024floquet,shi2025ultracritical}.

\medskip

\textit{\color{blue}The DC Nonlinear Conductivity.} By applying the microscopic formalism to a generic multiband system coupled to the fermionic bath, we compute the second-order DC conductivity $\sigma_{abc}$~\cite{supplemental}.
This result is derived directly from the exact NESS density matrix, obtained by solving the open-system dynamics without resorting to phenomenological regularizations. 
This microscopic derivation yields the precise analytical structure of the NESS, which dictates the subsequent expansion of the resulting conductivity by powers of $1/\Gamma$:
\begin{align}
\begin{aligned}
\sigma_{abc} = \frac{1}{\Gamma^2}\sigma_{abc}^{(-2)} + \frac{1}{\Gamma}\sigma_{abc}^{(-1)} +
\sigma_{abc}^{(0)} + {\cal O}(\Gamma),
\end{aligned}
\end{align}
We present the results using the Fermi-Dirac distribution derivatives $f_{0,n}^{(k)} = \partial^k f_{0} (\epsilon)/\partial \epsilon^k|_{\epsilon_n}$ (with $f'_{0, n} \equiv f^{(1)}_{0, n}$ following literature conventions), along with the conventional definition of the quantum metric $g_n^{ab} = \sum_{m\neq n}\text{Re}[A_{nm}^a A_{mn}^b]$~\cite{QGT_Provost_1980, QGT_PATI_1991}, and Berry curvature $\Omega^n_{ab} = -2\sum_{m\neq n}\text{Im}[A_{nm}^a A_{mn}^b]$~\cite{berry1984quantal, thouless1982quantized}, where $A^{a}_{nm} = \braket{n|i\partial_{a}|m}$ represents the Berry connection and $\partial_{a} \equiv \partial/\partial k_{a}$.

(1) $\Gamma$-dependent contributions ($\Gamma^{-2}, \Gamma^{-1}$): The leading divergence, $\sigma_{abc}^{(-2)}$, corresponds to the nonlinear Drude response:
\begin{equation}
\sigma_{abc}^{(-2)} = -\frac{1}{4} \sum_{n,{\bf k}} \partial_a(v_n^b v_n^c) f_{0,n}', 
\label{eq:sigma_neg2}
\end{equation}
While the ${\cal O}(\Gamma^{-1})$ term corresponds to the nonlinear Hall effect (NAHE), related to the Berry curvature $\Omega^n_{ab}$,
\begin{equation}
\sigma_{abc}^{(-1)} = \frac{1}{2} \sum_{n,{\bf k}} (v_n^c \Omega_{ab}^n + v_n^b \Omega_{ac}^n) f_{0,n}'.
\label{eq:sigma_neg1}
\end{equation}
This recovers the well-known expression for the Berry curvature dipole (BCD) contribution~\cite{sodemann2015quantum,ma2019observation,shi2023berry}, serving as a validation of the formalism.

(2) $\Gamma^0$ contributions: The $\Gamma$-independent term, $\sigma_{abc}^{(0)}$, is the central result of this work and the focus of discrepancies in the literature. 
The open-system formalism reveals a decomposition into distinct kinetic and geometric components: 
\begin{equation}
\sigma_{abc}^{(0)} = \sigma_{abc}^{\text{kin}} + \sigma_{abc}^{\text{geo}} .
\end{equation}

First, we identify a novel purely kinetic contribution:
\begin{equation}
\sigma^{\text{kin}}_{abc} = \sum_{n,{\bf k}} \frac{1}{24} v_n^a v_n^b v_n^c f_{0,n}^{(4)}.
\label{eq:sigma_kinetic}
\end{equation}
This term represents a $\Gamma$-independent, ballistic nonlinear response.
Its emergence is a direct consequence of the specific NESS regularization imposed by the fermionic bath, which leads to finite, $\Gamma^0$ corrections to the NESS occupation functions persisting in the clean limit ($\Gamma\to 0$). 
These corrections manifest analytically through high-order Polygamma functions~\cite{supplemental}, which generate the high-order energy derivative $f^{(4)}_0$, even in single-band models without Berry phases~\cite{matsyshyn2023fermi}. 
This regularization is physically distinct from that of RTAs, explaining why $\sigma_{abc}^{\text{kin}}$ is absent in those models.
This contribution is maximized at low temperatures or when sharp band structure features exist near the Fermi level.

Second, we obtain the contributions arising from the quantum geometry:
\begin{equation}
\sigma^{\text{geo}}_{abc} = \sum_{n,{\bf k}} \left[ (G_n^{ab} v_n^c + G_n^{ac} v_n^b - 2 G_n^{bc} v_n^a) -\frac{\partial_a g_n^{bc}}{2} \right] 2 f_{0,n}'.
\label{eq:sigma_geometric}
\end{equation}
This expression provides the definitive quantum metric contribution for a system coupled to the featureless fermionic bath. 
It separates into two physically distinct parts, both originating from field-induced geometric shifts of the electron wave packets.
The first term represents the energy-normalized interband quantum metric $G^{ab}_{n} \equiv \sum_{m\neq n}\text{Re}[A^{a}_{nm}A^{b}_{mn}]/(\epsilon_{n} - \epsilon_m)$~\cite{gao2014field, das2023intrinsic, kaplan2024unification}.
The second term, $-\partial_a g_{bc}/2$ is the intraband quantum metric contribution~\cite{ulrich2025quantum}.
Physically, it describes the contribution to the current arising from the deformation of the wave packet characterized by the metric $g_{bc}$, which quantifies the wave packet's spatial spread. 
It emerges explicitly when geometric contributions involving $f''$ are transformed to the Fermi surface ($f'$) via integration by parts~\cite{supplemental}.

The results summarized in Eqs.~(\ref{eq:sigma_kinetic}) and (\ref{eq:sigma_geometric}) provide the definitive analytical proof that the $\Gamma^0$ conductivity is non-universal. 
Both the kinetic contribution $\sigma_{a b c}^{\text{kin}}$ and the specific structure of $\sigma_{a b c}^{\text{geo}}$, including the intraband term $(-\partial_a g_{b c} / 2)$ and the exact prefactors of the interband terms, are characteristic of the fermionic bath. 
The discrepancies with RTA models, which exhibit varying geometric prefactors and lack the kinetic term (Table~\ref{tab:comparison}), demonstrate that $\sigma_{a b c}^{(0)}$ cannot be viewed as a property of the isolated system.

\medskip

\begin{figure}
\includegraphics[width=\columnwidth]{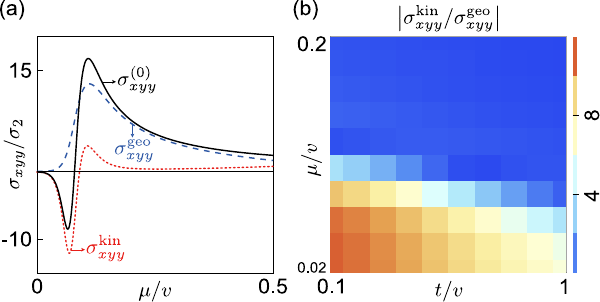}
\caption{Competition between $\Gamma^0$ geometric and kinetic nonlinear conductivity in the ${\cal P}{\cal T}$-symmetric model [Eq.~\eqref{eq:model_H}].
(a) Calculated $\Gamma^0$ contributions $\sigma_{xyy}^{\text{geo}}$ (blue dashed) and $\sigma_{xyy}^{\text{kin}}$ (red dotted), along with the total $\Gamma^0$ response $\sigma_{xyy}^{(0)}$ (black solid), as a function of chemical potential $\mu$. 
The kinetic term is comparable in magnitude to the geometric term. 
[Parameters used: $b/v=1$, $m/v=-3.9$, $t/v=0.9$, $\beta/v = 100$; characteristic second order conductivity $\sigma_2 = a (\hbar / v) \cdot e^3 / \hbar^2$].
(b) Color map of the ratio $|\sigma_{xyy}^{\text{kin}} / \sigma_{xyy}^{\text{geo}}|$ as a function of tilt parameter $t$ and chemical potential $\mu$. 
The kinetic term [Eq.~(\ref{eq:sigma_kinetic})] dominates (red regions) over the geometric term [Eq.~(\ref{eq:sigma_geometric})] in significant portions of the parameter space, underscoring its non-negligible role in the $\Gamma^0$ nonlinear response when the system is coupled to a fermionic bath.
}
\label{fig:model_calc}
\end{figure}

\medskip

\textit{\color{blue}Illustrative Model Calculation.} To demonstrate our central findings, the non-universality of $\sigma_{abc}^{(0)}$ and the physical significance of the novel kinetic term $\sigma_{abc}^{\text{kin}}$, we apply our microscopic formalism to a concrete lattice model \cite{qiang2025clarification}. 
We consider a 2D ${\cal P}{\cal T}$-symmetric metal on a square lattice (lattice constant $a$) described by the $4 \times 4$ tight-binding Hamiltonian:
\begin{align}
{\cal H} ({\bf k}) = v \tau_x\left[ \sin(k_x a) \sigma_x + \sin(k_y a) \sigma_y \right] + M({\bf k}) \tau_z,
\label{eq:model_H}
\end{align}
where $M({\bf k}) = m + b [ \cos(k_x a) + \cos(k_y a) ] + t \sin(k_x a)$.
This Hamiltonian is invariant under the combined anti-unitary symmetry ${\cal P}{\cal T}$ (with ${\cal P} = \tau_z$, ${\cal T} = i \sigma_y {\cal K}$) but explicitly breaks both the inversion symmetry ${\cal P}$ and the time-reversal symmetry ${\cal T}$ via the $t \sin(k_x a)$ term, making it an ideal testbed for second-order DC responses without BCD contributions \cite{gao2014field, liu2021intrinsic}.

The Hamiltonian [Eq.~\eqref{eq:model_H}] commutes with the symmetry operator $S = \tau_z \sigma_z$. 
This allows for a unitary transformation that separates ${\cal H}({\bf k})$ into the $S = \pm 1$ eigenspaces, yielding the block-diagonal form:
\begin{align}
{\cal H}'({\bf k}) = \begin{bmatrix}
h({\bf k}) & 0 \\
0 & h^*({\bf k})
\end{bmatrix},
\quad
h({\bf k}) = \begin{pmatrix}
M_{{\bf k}} & v_- \\
v_+ & -M_{{\bf k}}
\end{pmatrix} ,
\end{align}
with $v_\pm = v [\sin(k_x a) \pm i  \sin(k_y a)]$.
This structure simplifies the analysis by reducing the original $4 \times 4$ Hamiltonian to two copies of two-band system. 
We numerically evaluate the two competing $\Gamma^0$ contributions for this model: the novel kinetic term $\sigma_{xyy}^{\text{kin}}$ [Eq.~(\ref{eq:sigma_kinetic})] and the geometric term $\sigma_{xyy}^{\text{geo}}$ [Eq.~(\ref{eq:sigma_geometric})].
The results are presented in Fig.~\ref{fig:model_calc}.

Fig.~\ref{fig:model_calc}(a) compares the chemical potential ($\mu$) dependence of $\sigma_{xyy}^{\text{kin}}$ and $\sigma_{xyy}^{\text{geo}}$ for fixed model parameters. 
The geometric term, scaling as $\sim f'_{0}$, is a Fermi surface contribution that peaks near the band edges. 
The kinetic term, $\sigma_{xyy}^{\text{kin}} \propto f^{(4)}_{0}$, exhibits a more complex, oscillatory dependence on $\mu$. 
Due to the high-order derivative, $\sigma_{xyy}^{\text{kin}}$ is strongly sensitive to the sharpness of the band features near the Fermi level and is enhanced at low temperatures.

Fig.~\ref{fig:model_calc}(b) maps the ratio of the two contributions, $|\sigma_{xyy}^{\text{kin}} / \sigma_{xyy}^{\text{geo}}|$, as a function of the tilt parameter $t$ and chemical potential $\mu$. This demonstrates that $\sigma_{xyy}^{\text{kin}}$ is not merely a minor correction; it becomes comparable to, or even dominant over, the geometric term across large regions of the parameter space, particularly where the band curvature is significant.

This calculation demonstrates the quantitative significance of the mechanism-dependent kinetic term. 
While theories based on RTAs predict a {\it purely} geometric $\Gamma^0$ response (see Table.~\ref{tab:comparison}), the open-system analysis shows that the fermionic bath establishes a NESS where $\sigma_{a b c}^{\text{kin}}$ is a necessary and often comparable component of the total conductivity.

\medskip

\textit{\color{blue}Discussion and Conclusion.} We have established that the ``intrinsic'' $\Gamma^0$ nonlinear conductivity depends on the dissipation {\it mechanism}. 
By analyzing a Bloch system coupled to a microscopic fermionic bath, we derived an exact result for the second-order DC conductivity, $\sigma_{a b c}$, providing two key insights.

First, we clarified the structure of the $\Gamma^0$ geometric contribution, $\sigma_{a b c}^{\text{geo}}$ [Eq.~\eqref{eq:sigma_geometric}], confirming the presence of both intraband $(\sim \partial_a g_{b c})$ and specific interband terms. 
This result agrees with constant-$\Gamma$ Green's function formalisms~\cite{ulrich2025quantum}, suggesting $\sigma_{a b c}^{\text{geo}}$ may be characteristic for idealized Markovian models. 
The contrast with the conflicting results from RTA-based theories~\cite{das2023intrinsic,kaplan2024unification,qiang2025clarification} evidences that the dissipation mechanism shapes the quantum geometric response.

Second, we uncovered a novel, $\Gamma^0$ kinetic contribution, $\sigma_{a b c}^{\text {kin }} \sim v^3 f_0^{(4)}$ [Eq.~\eqref{eq:sigma_kinetic}]. 
This term emerges from the specific way the fermionic bath modifies the electronic occupation functions, which persists even in the clean limit.
The emergence of $\sigma_{a b c}^{\text{kin}}$ as a finite, mechanism-specific response, distinct from simplified phenomenological models, evidences this dependence. 
In fact, this dependence on the dissipation mechanism extends beyond DC responses and quantum geometry. 
For example, analysis of AC-driven Floquet systems shows that the rectification current differs when coupled to a bosonic versus a fermionic bath, even in a single Bloch band without quantum geometry and in the weak-coupling limit~\cite{shi2025ultracritical}.

Our open-system formulation of the dissipation-shaped NESS offers a new perspective on recent experiments and theories regarding the nonlinear responses in ${\cal P} {\cal T}$ symmetric antiferromagnets, specifically even-layered ${\rm MnBi}_2 {\rm Te}_4$\cite{gao2024antiferromagnetic, wang2023quantum, gao2023quantum}. These experiments extracted a $\Gamma$-independent intercept via scaling laws, operationally defining it as the intrinsic nonlinear response. However, Wang et al. \cite{wang2023quantum} measured a finite longitudinal response in pristine devices on bare substrates, while Gao et al. \cite{gao2023quantum,gao2024antiferromagnetic} reported a predominantly transverse response using encapsulated van der Waals heterostructures. Our findings offer a physical approach to reconcile these distinct observations by examining both the tensor structure of the response and the underlying dissipation mechanisms.

Analytically, the geometric response is not strictly restricted to be transverse. Within phenomenological RTA models \cite{qiang2025clarification,xiao2025proper}, the longitudinal geometric response vanishes due to the strict cancellation $(G^{a a} v^a+ G^{a a} v^a-2 G^{a a} v^a=0)$. 
However, in agreement with Ulrich et al.~\cite{ulrich2025quantum}, our geometric response shaped by the fermionic bath contains an intraband quantum metric term, $- \partial_a g_{b c} / 2$ [Eq.~\eqref{eq:sigma_geometric}].
Because this yields a non-zero term for the longitudinal direction $(- \partial_a g_{a a} / 2)$, it demonstrates that an intrinsic longitudinal geometric response is physically allowed.

This emergence of the intraband metric term is just one manifestation of a broader principle: the operationally extracted $\Gamma^0$ nonlinear response is not a universal band structure property, but is actively shaped by the specific dissipation mechanism. Extending beyond the geometric tensor structure, solving the exact dynamics for a system coupled to a wide-band fermionic bath shows the resulting nonlinear dc conductivity also includes a kinetic contribution [Eq.~\eqref{eq:sigma_kinetic}]. Because the aforementioned experiments utilized different architectures, ranging from bare substrates \cite{wang2023quantum} to encapsulated heterostructures \cite{gao2023quantum,gao2024antiferromagnetic}, the electrons coupled to distinct baths. Consequently, the variations in the extracted longitudinal-to-transverse ratios are physically expected: different dissipation mechanisms drive the system to distinct NESS, thereby generating different observable nonlinear transport signatures.

This principle of non-universality also clarifies ongoing theoretical efforts to define the intrinsic nonlinear current \cite{xiao2025proper, resta2025intrinsic}. Specifically, Xiao et al. \cite{xiao2025proper} demonstrated that a DC electric field shifts the local wavepacket energy, driving the system to a modified steady-state distribution. We support this overall physical picture but emphasize that the precise distribution is non-universal. Separately, Resta \cite{resta2025intrinsic} formulated the geometric positional shift within an exact many-body framework for an isolated system. Moving beyond this coherent limit, our approach emphasizes the importance of specific dissipation mechanisms in actively shaping the observable steady-state DC response.

This work reframes the understanding of ``intrinsic'' nonlinear response, emphasizing a departure from the linear response paradigm. 
The environment is not merely parameterized by $\Gamma$;
the specific dissipation {\it mechanism} actively shapes the non-equilibrium steady state density matrix, determining the precise occupation functions and coherences, and consequently, the observable conductivity.
The resulting non-universality indicates that a single, universal expression for the ``intrinsic'' conductivity is untenable. 
By showing that quantum geometry in nonlinear transport is shaped by dissipation, this work invites investigation into how specific, experimentally relevant physical environments influence nonlinear transport signatures.

\medskip

\textit{\color{blue}Acknowledgements.}
We would like to thank Yannis Ulrich for helpful discussions.
This work is supported by the National Natural Science Foundation of China (NSFC; Grants No. 12488101, and No. 92265203), the Strategic Priority Research Program of the Chinese Academy of Sciences (Grants No. XDB0460000 and No. XDB28000000), and the Quantum Science and Technology-National Science and Technology Major Project (Grants No. 2024ZD0300104 and No. 2021ZD0302600). H. W. acknowledges the support from the NSFC under Grants No. 12522411 and No. 12474240.

\bibliography{reference}

\clearpage

\newpage

\appendix

\renewcommand{\theequation}{\thesection-\arabic{equation}}
\renewcommand{\thefigure}{\thesection-\arabic{figure}}
\renewcommand{\thetable}{\thesection-\Roman{table}}

\onecolumngrid

\section*{SUPPLEMENTARY MATERIAL FOR: ``Dissipation-Shaped Quantum Geometry in Nonlinear Transport''}

This Supplementary Material provides a detailed derivation of the second-order DC nonlinear conductivity $\sigma_{abc}$ for an electronic system coupled to a microscopic fermionic heat bath. We demonstrate the derivation of the non-equilibrium steady state (NESS) density matrix and its expansion to obtain the conductivity components presented in the main text.

\section{Microscopic Model and NESS Density Matrix}
\label{S-A}

To establish a benchmark for the nonlinear conductivity, we derive the non-equilibrium steady state (NESS) density matrix from a microscopic open quantum system model. We consider a crystalline electronic system, $H_S(t)$, coupled to a non-interacting, wide-band fermionic bath, $H_B$, which models a metallic backgate.

The total Hamiltonian is $H(t) = H_S(t) + H_B + H_{SB}$, with components given by:
\begin{align}
H (t) = \begin{bmatrix}
H_S (t) & H_{SB} \\
H_{SB}^\dagger & H_B
\end{bmatrix} .
\end{align}
The system Hamiltonian, $H_S(t) = H_0 + V(t)$, describes Bloch electrons ($H_0$) subject to a general time-dependent perturbation $V(t)$. $H_0 = \sum_n \epsilon_n |\chi_n\rangle\langle \chi_n|$, where $|\chi_n\rangle$ and $\epsilon_n$ are the unperturbed eigenstates and energies, respectively. The indices $m, n, l$ hereafter are general, incorporating momentum, band, and spin degrees of freedom. This model is in the same class of those non-interacting fermionic models often described within the Keldysh formalism~\cite{nagaosa2017concept,gerchikov1989theory,fregoso2013driven,kamenev2011field,johnsen1999quasienergy,jauho1994time,kohler2005driven,matsyshyn2021rabi}.

The environment is modeled as a featureless fermionic bath, $H_B = \sum_{n, i} \varepsilon_i |\varphi_{n,i}\rangle\langle \varphi_{n,i}|$, where each system state $|\chi_n\rangle$ couples independently to a set of bath states $|\varphi_{n,i}\rangle$. The tunnel coupling is $H_{SB} = \lambda \sum_{n, i} (|\chi_{n}\rangle\langle \varphi_{n, i} | + \text{h.c.})$. We assume the bath is initially in thermal equilibrium, described by the Fermi-Dirac distribution $f_0(\epsilon)$ at chemical potential $\mu$ and temperature $T=1/\beta$.

Tracing out the bath degrees of freedom under the wide-band approximation (constant bath density of states, $\nu_B(\omega) = \nu_0$) yields the dynamics for the system's reduced density matrix $\rho_S(t)$. This procedure defines the physical relaxation rate $\Gamma = \lambda^2 \nu_0/2$, which microscopically regularizes the system dynamics. 

We solve for the NESS density matrix by expanding $\rho_S(t)$ perturbatively in the driving perturbation $V(t)$:
\begin{align}
\rho_S(t) = \rho^{(0)} + \rho^{(1)}(t) + \rho^{(2)}(t) + {\cal O}(V^3).
\end{align}
The zeroth-order term, or equilibrium NESS, is diagonal:
\begin{align}
\rho_{m n}^{(0)} = \delta_{m n}
\int_{-\infty}^{\infty} \frac{\text{d} \omega_b}{2 \pi}
\frac{ 2 \Gamma }
{\omega_b^2 + \Gamma^2}
f_0(\epsilon_{m} + \omega_b) \equiv \delta_{mn} \tilde{\rho}_n^{(0)}.
\label{eq:SM_rho0}
\end{align}
This expression represents the equilibrium Fermi-Dirac distribution, broadened by the system-bath coupling $\Gamma$. In the ideal-bath limit ($\Gamma \to 0$), $\tilde{\rho}_n^{(0)} \to f_0(\epsilon_n)$.

The first- and second-order corrections to the density matrix define the response kernels. We express them in the frequency domain, where $V(t) = \int V(\omega) e^{-i\omega t} d\omega / (2\pi)$ and $\rho^{(n)}(t) = \int \rho^{(n)}(\omega) e^{-i\omega t} d\omega / (2\pi)$. The response kernels $\tilde{\rho}^{(1)}$ and $\tilde{\rho}^{(2)}$ are found to be:
\begin{align}
\rho_{m n}^{(1)} (\omega) = V_{m n} (\omega) \, \tilde{\rho}_{m n}^{(1)} (\omega) ,
\quad
\tilde{\rho}_{m n}^{(1)} (\omega) = \int_{-\infty}^{\infty} \frac{\text{d} \omega_b}{2 \pi} \frac{ 2\Gamma } {\omega_b^2 + \Gamma^2}
\frac{ f_0(\epsilon_{n} + \omega_b ) - f_0(\epsilon_{m} - \omega_b )}
{ \omega + \omega_b + \epsilon_{n m} + i \Gamma } ,
\label{eq:SM_rho1}
\end{align}
and
\begin{align}
& \rho_{m n}^{(2)} (\omega_1 + \omega_2) = \sum_{l} V_{m l} (\omega_1) V_{l n} (\omega_2)
\tilde{\rho}_{m l n}^{(2)} (\omega_1, \omega_2) ,
\nonumber \\
& \tilde{\rho}_{m l n}^{(2)} (\omega_1,\omega_2) = \int_{-\infty}^{\infty} \frac{\text{d} \omega_b}{2 \pi} \frac{ 2 \Gamma }
{\omega_b^2 + \Gamma^2} \bigg[
\frac{f_0(\epsilon_{m} - \omega_b )}{ (\omega_1 + \omega_2 + \omega_b + \epsilon_{n m} + i \Gamma) (\omega_1 + \omega_b + \epsilon_{l m} + i \Gamma )}
\nonumber \\
& \quad + \frac{f_0(\epsilon_{n} + \omega_b )}{ ( \omega_1 + \omega_2 + \omega_b + \epsilon_{n m} + i \Gamma ) (\omega_2 + \omega_b + \epsilon_{n l} + i \Gamma )}
- \frac{ f_0(\epsilon_{l} + \omega_b ) }{ ( \omega_1 + \omega_b + \epsilon_{l m} + i \Gamma ) ( \omega_2 - \omega_b + \epsilon_{n l} + i \Gamma ) }
\bigg] ,
\label{eq:SM_rho2}
\end{align}
where $\epsilon_{nm} \equiv \epsilon_n - \epsilon_m$. These expressions, derived directly from the open-system dynamics, provide the exact response kernels for a system coupled to a fermionic bath. They differ from phenomenological models (e.g., relaxation time approximations and imaginary frequency regularizations) and form the basis for calculating the nonlinear conductivity.

\medskip
\subsection*{Alternative Representation}
\label{S-AA}

The integrals over the bath variable $\omega_b$ in Eqs. (\ref{eq:SM_rho0}) to (\ref{eq:SM_rho2}) can be performed using the Cauchy residue theorem. This leads to an alternative representation in terms of the Polygamma function $\psi^{(0)}$.  We find the broadened zero-th kernel
\begin{align}
\rho_{m n}^{(0)} = \delta_{m n}
\frac{1}{2}
\Big[
f_+ (\epsilon_n) + f_- (\epsilon_n)
\Big] \equiv \tilde{\rho}_n^{(0)},
\quad
f_\pm (\epsilon) = \frac{1}{2} \pm \frac{i}{\pi}\psi^{(0)}
\left(\frac{1}{2} \pm i\beta \frac{\epsilon \mp i \Gamma -\mu}{2\pi}\right),
\label{rho_0_supp}
\end{align}
where $\psi^{(0)}$ is the 0-th order Polygamma function.
Importantly,
\begin{align}
\lim_{\Gamma \to 0} \tilde{\rho}_n^{(0)} = \frac{1}{1+\exp[\beta (\epsilon_n - \mu)]} ,
\end{align}
which shows that $\tilde{\rho}_n^{(0)}$ reduces to the ideal Fermi-Dirac distribution in the limit of $\Gamma \to 0$.
Moreover, for the first- and second-order expansions, we have 
\begin{align}
& \tilde{\rho}_{m n}^{(1)} (\omega) = 
\frac{ \tilde{\rho}_n^{(0)} - \tilde{\rho}_m^{(0)} }
{ \omega + \epsilon_{n m}  + 2 i \Gamma }
+
\frac{r_+^{(1)} (\omega, \epsilon_m, \epsilon_n) + r_-^{(1)} (\omega, \epsilon_m, \epsilon_n) }
{ \omega + \epsilon_{n m}  + 2 i \Gamma },
\label{tilde-rho_1_supp}
\end{align}
with
\begin{align}
r_+^{(1)} (\omega, \epsilon_m, \epsilon_n) 
= i \Gamma \frac{ f_+ (\epsilon_n) -  f_+ (\epsilon_m - \omega) }{\omega + \epsilon_{n m} },
\quad
r_-^{(1)} (\omega, \epsilon_m, \epsilon_n) 
= i \Gamma \frac{ f_- (\epsilon_n + \omega ) -  f_- (\epsilon_m) }{\omega + \epsilon_{n m} },
\end{align}
and
\begin{align}
& \tilde{\rho}_{m l n}^{(2)} (\omega_1,\omega_2) =  
\frac{ 
\tilde{\rho}_{l n}^{(1)} (\omega_2) - 
\tilde{\rho}_{m l}^{(1)} (\omega_1) 
}
{ \omega_1 + \omega_2 + \epsilon_{n m}  + 2 i \Gamma }
+
\frac{ 
r_+^{(2)}  (\omega_1, \omega_2, \epsilon_m, \epsilon_l, \epsilon_n) 
+ r_-^{(2)}  (\omega_1, \omega_2, \epsilon_m, \epsilon_l, \epsilon_n) }
{ \omega_1 + \omega_2 + \epsilon_{n m}  + 2 i \Gamma },
\end{align}
in which
\begin{align}
r_+^{(2)}  (\omega_1, \omega_2, \epsilon_m, \epsilon_l, \epsilon_n)
= \frac{ 
r_+^{(1)} (\omega_2,\epsilon_l,\epsilon_n)  
- r_+^{(1)} (\omega_1, \epsilon_m - \omega_2, \epsilon_l - \omega_2)  }
{\omega_1 + \omega_2 + \epsilon_{n m} },
\end{align}
\begin{align}
r_-^{(2)}  (\omega_1, \omega_2, \epsilon_m, \epsilon_l, \epsilon_n) 
= \frac{ 
r_-^{(1)} (\omega_2, \epsilon_l+\omega_1, \epsilon_n+\omega_1)  
- r_-^{(1)} (\omega_1, \epsilon_m, \epsilon_l)  }
{\omega_1 + \omega_2 + \epsilon_{n m} }.
\label{tilde-rho_2_supp}
\end{align}
From Eqs.~(\ref{tilde-rho_1_supp}) and (\ref{tilde-rho_2_supp}), it is clear that the perturbative expansions of the density matrix for the system coupled with the featureless fermionic bath, obtained by solving the open system Schr\"{o}dinger equation exactly, have striking differences with those obtained from conventional perturbation theories assuming an adiabatic turning-on (IFRs) or a hand-given relaxation (RTAs).

\section{General Expression for Nonlinear Conductivity}
\label{S-B}

We now apply the microscopic NESS formalism to derive the general expression for the second-order nonlinear conductivity, $\sigma_{abc}$. The physical perturbation $V(t)$ is induced by a uniform electric field ${\bf E}(t)$ via the Peierls substitution, $H_S(t) = H_0({\bf k} - {\bf A}(t))$, where ${\bf E}(t) = -\partial_t {\bf A}(t)$.

The perturbation $V(t) = H_S(t) - H_0$ and the current operator $J_a(t) = \partial H_S(t) / \partial k_a$ are expanded in powers of the vector potential ${\bf A}(t)$. Using $\partial_a \equiv \partial_{k_a}$, we have:
\begin{align}
V(t) &= V^{(1)}(t) + V^{(2)}(t) + {\cal O}(A^3)
= \sum_b (-\partial_b H_0) A_b(t) + \sum_{b,c} \frac{1}{2} (\partial_b \partial_c H_0) A_b(t) A_c(t) + \dots
\\
J_a(t) &= J_a^{(0)} + J_a^{(1)}(t) + J_a^{(2)}(t) + {\cal O}(A^3)
= (\partial_a H_0) + \sum_c (-\partial_a \partial_c H_0) A_c(t) + \sum_{b,c} \frac{1}{2} (\partial_a \partial_b \partial_c H_0) A_b(t) A_c(t) + \dots
\end{align}
The total nonlinear current is the expectation value $j_a(t) = \text{Tr}[\rho_S(t) J_a(t)]$. The component at second order in the field, $j_a^{(2)}(t)$, receives contributions from all combinations of density matrix and current operator orders that multiply to ${\cal O}(A^2)$:
\begin{align}
j_a^{(2)}(t) = \text{Tr}\big[ \rho^{(0)} J_a^{(2)}(t) \big] + \text{Tr}\big[ \rho^{(1)}(t) J_a^{(1)}(t) \big] + \text{Tr}\big[ \rho^{(2)}(t) J_a^{(0)} \big].
\end{align}
It is important to correctly identify the sources for the density matrix terms:
\begin{itemize}
    \item $\rho^{(1)}(t)$ contains terms linear in $V$, so it has parts from $V^{(1)}$ and $V^{(2)}$: $\rho^{(1)}(t) = \rho^{(1)}[V^{(1)}] + \rho^{(1)}[V^{(2)}] + \dots$
    \item $\rho^{(2)}(t)$ contains terms quadratic in $V$. The ${\cal O}(A^2)$ term arises from $V^{(1)}V^{(1)}$: $\rho^{(2)}(t) = \rho^{(2)}[V^{(1)}, V^{(1)}] + \dots$
\end{itemize}
This leads to four distinct contributions to the second-order current:
\begin{align}
j_a^{(2)}(t) = \underbrace{\text{Tr}[\rho^{(0)} J_a^{(2)}(t)]}_{\text{(I)}}
+ \underbrace{\text{Tr}[\rho^{(1)}[V^{(1)}] J_a^{(1)}(t)]}_{\text{(II)}}
+ \underbrace{\text{Tr}[\rho^{(1)}[V^{(2)}] J_a^{(0)}]}_{\text{(III)}}
+ \underbrace{\text{Tr}[\rho^{(2)}[V^{(1)}, V^{(1)}] J_a^{(0)}]}_{\text{(IV)}}.
\end{align}
The conductivity tensor $\sigma_{abc}(\omega_1, \omega_2)$ is defined in frequency space via $j_a^{(2)}(\omega_1 + \omega_2) = \sigma_{abc}(\omega_1, \omega_2) E_b(\omega_1) E_c(\omega_2)$. Using $A_b(\omega) = E_b(\omega) / (i\omega)$ and the response kernels from Eqs. (\ref{eq:SM_rho1}) to (\ref{eq:SM_rho2}), we map each of the four terms to the conductivity.

We introduce the operator matrix elements for the Fourier components, corresponding to the notation in:
$V_{mn}^{(b)} \equiv \langle m | (-\partial_b H_0) | n \rangle$, $V_{mn}^{(bc)} \equiv \langle m | (\partial_b \partial_c H_0) | n \rangle / 2$, $J_{mn}^{(a)} \equiv \langle m | (\partial_a H_0) | n \rangle = - V_{mn}^{(a)}$, $J_{mn}^{(ac)} \equiv \langle m | (-\partial_a \partial_c H_0) | n \rangle = - V_{mn}^{(ac)} $, and $J_{mn}^{(abc)} \equiv \langle m | \partial_a \partial_b \partial_c H_0 | n \rangle / 2$.
Translating terms (I)-(IV) into the full frequency-dependent conductivity, we obtain the expression:
\begin{align}
\sigma_{abc} (\omega_1, \omega_2) = \frac{i}{\omega_1} \frac{i}{\omega_2} \sum_{m,l,n} \bigg[
& \underbrace{ V_{ml}^{(b)} V_{ln}^{(c)} \tilde{\rho}_{mln}^{(2)} (\omega_1,\omega_2) J_{nm}^{(a)} }_{\text{from (IV): } \rho^{(2)}[V^{(1)},V^{(1)}] J^{(0)}}
+ \underbrace{ V_{mn}^{(bc)} \tilde{\rho}_{mn}^{(1)} (\omega_1+\omega_2) J_{nm}^{(a)} }_{\text{from (III): } \rho^{(1)}[V^{(2)}] J^{(0)}}
\nonumber \\
+ & \underbrace{ V_{mn}^{(b)} \tilde{\rho}_{mn}^{(1)} (\omega_1) J_{nm}^{(ac)} }_{\text{from (II): } \rho^{(1)}[V^{(1)}] J^{(1)}}
+ \underbrace{ \rho_{mn}^{(0)} J_{nm}^{(abc)} }_{\text{from (I): } \rho^{(0)} J^{(2)}}
\bigg]
+ \Big(
\begin{matrix}
b \leftrightarrow c
\\
\omega_1 \leftrightarrow \omega_2
\end{matrix}
\Big).
\label{eq:SM_sigma_full}
\end{align}
This expression is the exact second-order conductivity for the microscopic model. The first term, $\rho_{mn}^{(0)} J_{nm}^{(abc)}$, is the ${\cal O}(A^2)$ correction to the current operator itself (term I). The second term, involving $J_{a,nm}^{(c)}$, arises from the ${\cal O}(A)$ correction to the density matrix and the ${\cal O}(A)$ correction to the current operator (term II). The final two terms, involving $J_{nm}^{(a)}$, arise from the second-order correction to the density matrix, which has a component from the first-order response to the ${\cal O}(A^2)$ perturbation (term III, $V^{(2)}$) and a component from the second-order response to the ${\cal O}(A)$ perturbation (term IV, $V^{(1)}V^{(1)}$).

The DC conductivity, 
\begin{align}
\sigma_{abc} \equiv \lim_{\omega_1 \to 0}
\Big[\lim_{\omega_2 \to -\omega_1} \sigma_{abc}(\omega_1, \omega_2) \Big] ,
\end{align}
is obtained by taking two sequential limits of Eq. (\ref{eq:SM_sigma_full}). The specific structure of the NESS kernels $\tilde{\rho}^{(n)}$, particularly their regularization by $\Gamma$, decides the final form of the intrinsic ($\Gamma$-independent) and extrinsic ($\Gamma$-dependent) contributions.

\section{Structure of Second-Order Expansion Coefficients}
\label{S-C}

In this section, we provide the detailed analytical expressions for the second-order expansion coefficients of the density matrix response kernels, $\tilde{\rho}_{m n}^{(1)} (\omega)$ and $\tilde{\rho}_{m l n}^{(2)} (\omega, \omega_2 \to - \omega)$, derived from the microscopic open-system formalism for a generic multiband model. These coefficients are essential for the regularization of the DC limit and the calculation of the nonlinear conductivity.

We define the expansion coefficients ${\cal C}_{mn}^{(1, k)}$ and ${\cal C}_{mln}^{(2, k)}$ (for the specific frequency configuration $\omega_1=\omega, \omega_2=-\omega$) via Taylor expansion around $\omega=0$:
\begin{align}
\tilde{\rho}_{m n}^{(1)} (\omega) = \sum_{k=0}^{\infty} {\cal C}_{mn}^{(1, k)} \, \omega^k, \quad
\tilde{\rho}_{m l n}^{(2)} (\omega, -\omega) = \sum_{k=0}^{\infty} {\cal C}_{mln}^{(2, k)} \, \omega^k.
\end{align}
We verified analytically that the current $\lim_{\omega' \to -\omega}j_a^{(2)}(\omega + \omega')$ from the zeroth-order and first-order coefficients ($k=0,1$) are exactly zero, and present the results for the second-order coefficients ($k=2$). The expressions are given in terms of the Polygamma functions $\psi^{(n)}(z)$. We utilize the shorthand notation $\epsilon_{nm} = \epsilon_n - \epsilon_m$ for the energy difference and define the arguments of the Polygamma functions as:
\begin{align}
\begin{aligned}
z_{n, \pm} = \frac{1}{2} + \frac{\beta}{2\pi}(\Gamma \pm i(\epsilon_n - \mu)) \quad
z_{n, \pm}^0 = \frac{1}{2} \pm \frac{i \beta}{2\pi} (\epsilon_n - \mu)
\end{aligned}
\end{align}
where $\beta$ is the inverse temperature, $\Gamma$ is the bath-induced relaxation rate, and $\mu$ is the chemical potential.

\subsection{First-Order Kernel Coefficients}

\subsubsection{Intraband Coefficient: \texorpdfstring{${\cal C}_{nn}^{(1, 2)}$}{Cnn}}
\begin{align}
{\cal C}_{nn}^{(1, 2)} = \frac{\beta}{96 \pi^4 \Gamma^2} \bigg[ & 6 \pi^2 \left(\psi^{(1)}(z_{n,+}) + \psi^{(1)}(z_{n,-})\right) - 3 \pi \beta \Gamma \left(\psi^{(2)}(z_{n,+}) + \psi^{(2)}(z_{n,-})\right) 
\nonumber \\
& + (\beta \Gamma)^2 \left(\psi^{(3)}(z_{n,+}) + \psi^{(3)}(z_{n,-})\right) \bigg]
\end{align}
Its Taylor expansion in $\Gamma$ is:
\begin{align}
{\cal C}_{nn}^{(1, 2)} = & \frac{1}{\Gamma^2} \left[ \frac{\beta}{16\pi^2} \sum_{\pm} \psi^{(1)}(z_{n,\pm}^0) \right] + \left[ \frac{\beta^3}{384\pi^4} \sum_{\pm} \psi^{(3)}(z_{n,\pm}^0) \right] + {\cal O}(\Gamma)
\end{align}
where $\psi^{(n)}$ is the $n$-th order Polygamma function, which can be converted to $n$-th order derivative over the Fermi-Dirac function [see Eq.~\eqref{rho_0_supp}].

\subsubsection{Interband Coefficient: \texorpdfstring{${\cal C}_{nm}^{(1, 2)}$}{Cnm}}
\begin{align}
{\cal C}_{nm}^{(1, 2)} = \frac{i}{8 \pi^3 \epsilon_{nm}^3 (2 i\Gamma - \epsilon_{nm})^2} \Bigg[
& 4 \pi^2 (2 \Gamma + i \epsilon_{nm})^2 \left( - \psi^{(0)}(z_{n,+}) + \psi^{(0)}(z_{m,+}) + \psi^{(0)}(z_{n,-}) - \psi^{(0)}(z_{m,-}) \right) \nonumber \\
& + 8 \pi \beta \Gamma \epsilon_{nm} (i \Gamma - \epsilon_{nm}) \left( \psi^{(1)}(z_{n,+}) + \psi^{(1)}(z_{m,-}) \right) \nonumber \\
& + \beta^2 \Gamma \epsilon_{nm}^2 (2 \Gamma + i \epsilon_{nm}) \left( \psi^{(2)}(z_{n,+}) + \psi^{(2)}(z_{m,-}) \right)
\Bigg]
\end{align}
Its Taylor expansion in $\Gamma$ is:
\begin{align}
{\cal C}_{nm}^{(1, 2)} = -\frac{i}{2\pi \epsilon_{nm}^3} \left[ \psi^{(0)}(z_{n,-}^0) - \psi^{(0)}(z_{n,+}^0) - \psi^{(0)}(z_{m,-}^0) + \psi^{(0)}(z_{m,+}^0) \right] + {\cal O}(\Gamma)
\end{align}

\subsection{Second-Order Kernel Coefficients}

\subsubsection{Fully Intraband Coefficient: \texorpdfstring{${\cal C}_{nnn}^{(2, 2)}$}{Cnnn}}
\label{Fully-Intraband-Coefficient}
\begin{align}
{\cal C}_{nnn}^{(2, 2)} = \frac{i \beta^4}{768 \pi^5} \left[ \psi^{(4)}(z_{n,+}) - \psi^{(4)}(z_{n,-}) \right]
\end{align}
Its Taylor expansion in $\Gamma$ is:
\begin{align}
{\cal C}_{nnn}^{(2, 2)} = \frac{i \beta^4}{768 \pi^5} \left[ \psi^{(4)}(z_{n,+}^0) - \psi^{(4)}(z_{n,-}^0) \right] + {\cal O}(\Gamma)
\end{align}
The presence of the fourth-order Polygamma function $\psi^{(4)}$ is a direct consequence of the regularization imposed by the fermionic bath. This specific analytical structure is responsible for the emergence of the intrinsic kinetic contribution $\sigma_{a b c}^{\text{kin}} \propto f_0^{(4)}$ in the clean limit (see Section~\ref{A-Kinetic-Contribution}).

\subsubsection{Mixed Coefficient: \texorpdfstring{${\cal C}_{nmn}^{(2, 2)}$}{Cnmn}}
The expression for ${\cal C}_{nmn}^{(2, 2)}$ is composed of terms involving $\psi^{(0)}$, $\psi^{(1)}$, and $\psi^{(2)}$:
\begin{align}
{\cal C}_{nmn}^{(2, 2)} = P_0 + P_1 + P_2
\end{align}
where
\begin{align}
P_0 &= -\frac{6 i \pi}{4 \pi^2 \epsilon_{nm}^4} \left[ \psi^{(0)}(z_{n,+}) - \psi^{(0)}(z_{m,+}) - \psi^{(0)}(z_{n,-}) + \psi^{(0)}(z_{m,-}) \right]
\end{align}
\begin{align}
P_1 = \frac{\beta \epsilon_{nm}}{(4 \Gamma^2 + \epsilon_{nm}^2)^2} \Bigg[
& (4 \Gamma^2 - 2 i \Gamma \epsilon_{nm} + \epsilon_{nm}^2) (2 i \Gamma - \epsilon_{nm})^2 \psi^{(1)}(z_{n,-}) \nonumber \\
& + (4 \Gamma^2 + 2 i \Gamma \epsilon_{nm} + \epsilon_{nm}^2) (2 i \Gamma + \epsilon_{nm})^2 \psi^{(1)}(z_{n,+}) \nonumber \\
& + 2 \Gamma (4 \Gamma + 3 i \epsilon_{nm}) (2 i \Gamma + \epsilon_{nm})^2 \psi^{(1)}(z_{m,-}) \nonumber \\
& - 2 i \Gamma (4 i \Gamma + 3 \epsilon_{nm}) (2 i \Gamma - \epsilon_{nm})^2 \psi^{(1)}(z_{m,+})
\Bigg]
\end{align}
and
\begin{align}
P_2 = - \frac{i \beta^2 \epsilon_{nm}^2}{4 \pi (4 \Gamma^2 + \epsilon_{nm}^2)} \Bigg[
& \epsilon_{nm} (2 i \Gamma + \epsilon_{nm}) \psi^{(2)}(z_{n,+})
+ 2 \Gamma (2 \Gamma + i \epsilon_{nm}) \psi^{(2)}(z_{m,+}) \nonumber \\
& + \epsilon_{nm} (2 i \Gamma - \epsilon_{nm}) \psi^{(2)}(z_{n,-})
- 2 \Gamma (2 \Gamma - i \epsilon_{nm}) \psi^{(2)}(z_{m,-})
\Bigg]
\end{align}
Its Taylor expansion in $\Gamma$ is:
\begin{align}
{\cal C}_{nmn}^{(2, 2)} = \bigg[ & \frac{3i}{2\pi\epsilon_{nm}^4} \left( \psi^{(0)}(z_{n,-}^0) - \psi^{(0)}(z_{n,+}^0) - \psi^{(0)}(z_{m,-}^0) + \psi^{(0)}(z_{m,+}^0) \right) \nonumber \\
& + \frac{\beta}{4\pi^2\epsilon_{nm}^3} \sum_{\pm} \psi^{(1)}(z_{n,\pm}^0) - \frac{i\beta^2}{16\pi^3\epsilon_{nm}^2} \left( \psi^{(2)}(z_{n,+}^0) - \psi^{(2)}(z_{n,-}^0) \right) \bigg] + {\cal O}(\Gamma)
\end{align}

\subsubsection{Mixed Coefficient: \texorpdfstring{${\cal C}_{nnm}^{(2, 2)}$}{Cnnm}}
The expression for ${\cal C}_{nnm}^{(2, 2)}$ involves terms up to $\psi^{(3)}$.
\begin{align}
{\cal C}_{nnm}^{(2, 2)} = \frac{1}{96 \pi^4 (2 i \Gamma - \epsilon_{nm})} \left[ Q_0 + Q_1 + Q_2 + Q_3 \right]
\end{align}
where the terms $Q_n$ correspond to contributions involving $\psi^{(n)}$:
\begin{align}
Q_0 = \frac{48 \pi^3}{\epsilon_{nm}^4}  \Bigg[
& -i (-2 i \Gamma + \epsilon_{nm}) \psi^{(0)}(z_{n,-})
+ (2 \Gamma + i \epsilon_{nm}) \psi^{(0)}(z_{n,+}) \nonumber \\
& + (2 \Gamma + i \epsilon_{nm}) \psi^{(0)}(z_{m,-})
- i (-2 i \Gamma + \epsilon_{nm}) \psi^{(0)}(z_{m,+})
\Bigg]
\end{align}
\begin{align}
Q_1 = 6 \pi^2 \beta \Bigg[
& \frac{1}{\Gamma^2} \left(-1 - \frac{8 i \Gamma^3}{\epsilon_{nm}^3}\right) \psi^{(1)}(z_{n,-}) - \frac{16 \Gamma (\Gamma + i \epsilon_{nm})}{\epsilon_{nm}^2 (2 i \Gamma - \epsilon_{nm})^2} \psi^{(1)}(z_{m,-})
\nonumber \\
& + \left(-\frac{1}{\Gamma^2} + \frac{4}{\epsilon_{nm}^2} - \frac{8 i \Gamma}{\epsilon_{nm}^3} - \frac{4}{(2 i \Gamma - \epsilon_{nm})^2}\right) \psi^{(1)}(z_{n,+}) 
\Bigg]
\end{align}
\begin{align}
Q_2 = 3 \pi \beta^2 \Bigg[
\frac{4 \Gamma^2 + \epsilon_{nm}^2}{\Gamma \epsilon_{nm}^2} \psi^{(2)}(z_{n,-}) + \frac{1}{\Gamma} \left(1 + \frac{8 \Gamma^2 (i \Gamma - \epsilon_{nm})}{\epsilon_{nm}^2 (-2 i \Gamma + \epsilon_{nm})}\right) \psi^{(2)}(z_{n,+}) + \frac{4 \Gamma}{\epsilon_{nm} (2 i \Gamma - \epsilon_{nm})} \psi^{(2)}(z_{m,-})
\Bigg]
\end{align}
\begin{align}
Q_3 = \frac{\beta^3 (2 i \Gamma - \epsilon_{nm})}{\epsilon_{nm}} \left[ \psi^{(3)}(z_{n,-}) + \psi^{(3)}(z_{n,+}) \right]
\end{align}
Its Taylor expansion in $\Gamma$ is:
\begin{align}
{\cal C}_{nnm}^{(2, 2)} = & \frac{1}{\Gamma^2} \left[ \frac{\beta}{16\pi^2\epsilon_{nm}} \sum_{\pm} \psi^{(1)}(z_{n,\pm}^0) \right] + \frac{1}{\Gamma} \left[ \frac{i\beta}{8\pi^2\epsilon_{nm}^2} \sum_{\pm} \psi^{(1)}(z_{n,\pm}^0) \right] \nonumber \\
& +  \bigg[ -\frac{\beta}{4\pi^2\epsilon_{nm}^3} \sum_{\pm} \psi^{(1)}(z_{n,\pm}^0) + \frac{\beta^3}{384\pi^4\epsilon_{nm}} \sum_{\pm} \psi^{(3)}(z_{n,\pm}^0) \nonumber \\
& \qquad + \frac{i}{2\pi\epsilon_{nm}^4} \left( \psi^{(0)}(z_{n,-}^0) - \psi^{(0)}(z_{n,+}^0) - \psi^{(0)}(z_{m,-}^0) + \psi^{(0)}(z_{m,+}^0) \right) \bigg] + {\cal O}(\Gamma)
\end{align}

\subsubsection{Mixed Coefficient: \texorpdfstring{${\cal C}_{nmm}^{(2, 2)}$}{Cnmm}}
Similarly, the expression for ${\cal C}_{nmm}^{(2, 2)}$ is given by:
\begin{align}
{\cal C}_{nmm}^{(2, 2)} = \frac{1}{96 \pi^4 (2 i \Gamma - \epsilon_{nm})} \left[ U_0 + U_1 + U_2 + U_3 \right]
\end{align}
where
\begin{align}
U_0 = \frac{48 \pi^3}{\epsilon_{nm}^4}  \Bigg[
& (2 \Gamma + i \epsilon_{nm}) \psi^{(0)}(z_{n,-})
- i (-2 i \Gamma + \epsilon_{nm}) \psi^{(0)}(z_{m,+}) \nonumber \\
& - i (-2 i \Gamma + \epsilon_{nm}) \psi^{(0)}(z_{m,-})
+ (2 \Gamma + i \epsilon_{nm}) \psi^{(0)}(z_{m,+})
\Bigg]
\end{align}
\begin{align}
U_1 = 6 \pi^2 \beta \Bigg[
& \frac{16 \Gamma (\Gamma + i \epsilon_{nm})}{\epsilon_{nm}^2 (2 i \Gamma - \epsilon_{nm})^2} \psi^{(1)}(z_{n,+}) + \frac{1}{\Gamma^2} \left(1 + \frac{8 i \Gamma^3}{\epsilon_{nm}^3}\right) \psi^{(1)}(z_{m,+}) \nonumber \\
& + \frac{1}{\Gamma^2} \left(1 + 4 \Gamma^2 \left(\frac{1}{(2 i \Gamma - \epsilon_{nm})^2} + \frac{2 i \Gamma - \epsilon_{nm}}{\epsilon_{nm}^3}\right)\right) \psi^{(1)}(z_{m,-}) 
\Bigg]
\end{align}
\begin{align}
U_2 = 3 \pi \beta^2 \Bigg[
& \frac{4 \Gamma}{\epsilon_{nm} (-2 i \Gamma + \epsilon_{nm})} \psi^{(2)}(z_{n,+}) - \frac{4 \Gamma^2 + \epsilon_{nm}^2}{\Gamma \epsilon_{nm}^2} \psi^{(2)}(z_{m,+}) \nonumber \\
& + \frac{1}{\Gamma} \left(-1 + \frac{4 \Gamma^2}{\epsilon_{nm}^2} \left(2 - \frac{2 i \Gamma}{2 i \Gamma - \epsilon_{nm}}\right)\right) \psi^{(2)}(z_{m,-})
\Bigg]
\end{align}
\begin{align}
U_3 = \frac{\beta^3 (-2 i \Gamma + \epsilon_{nm})}{\epsilon_{nm}} \left[ \psi^{(3)}(z_{m,-}) + \psi^{(3)}(z_{m,+}) \right]
\end{align}
Its Taylor expansion in $\Gamma$ is:
\begin{align}
{\cal C}_{nmm}^{(2, 2)} = & \frac{1}{\Gamma^2} \left[ -\frac{\beta}{16\pi^2\epsilon_{nm}} \sum_{\pm} \psi^{(1)}(z_{m,\pm}^0) \right] + \frac{1}{\Gamma} \left[ -\frac{i\beta}{8\pi^2\epsilon_{nm}^2} \sum_{\pm} \psi^{(1)}(z_{m,\pm}^0) \right] \nonumber \\
& + \bigg[ \frac{\beta}{4\pi^2\epsilon_{nm}^3} \sum_{\pm} \psi^{(1)}(z_{m,\pm}^0) - \frac{\beta^3}{384\pi^4\epsilon_{nm}} \sum_{\pm} \psi^{(3)}(z_{m,\pm}^0) \nonumber \\
& \qquad - \frac{i}{2\pi\epsilon_{nm}^4} \left( \psi^{(0)}(z_{n,-}^0) - \psi^{(0)}(z_{n,+}^0) - \psi^{(0)}(z_{m,-}^0) + \psi^{(0)}(z_{m,+}^0) \right) \bigg] + {\cal O}(\Gamma)
\end{align}

These expressions highlight the complex analytical structure of the response kernels, characterized by the interplay between the energy denominators ($\epsilon_{nm}$) and the relaxation rate ($\Gamma$), reflecting the specific regularization imposed by the fermionic bath.

\subsubsection{Fully Interband Coefficient: \texorpdfstring{${\cal C}_{nml}^{(2, 2)}$}{Cnml}}
The fully interband coefficient involves three distinct band indices, $n, m,$ and $l$, and comprises contributions from $\psi^{(0)}$, $\psi^{(1)}$, and $\psi^{(2)}$:
\begin{equation}
	{\cal C}_{nml}^{(2, 2)} = \frac{1}{2i\Gamma - \epsilon_{nm}}(R_0 + R_1 + R_2)
\end{equation}
where
\begin{equation}
	\begin{aligned}
		R_0 = & \psi^{(0)}(z_{n-}) \frac{2 \Gamma + i \epsilon_{nm}}{2 \pi \epsilon_{nm} \epsilon_{nl}^3} - \frac{i \psi^{(0)}(z_{n+}) (-2 i \Gamma + \epsilon_{nm})}{2 \pi \epsilon_{nm} \epsilon_{nl}^3}
		+ \psi^{(0)}(z_{m+}) \frac{2 \Gamma + i \epsilon_{nm}}{2 \pi \epsilon_{nm} \epsilon_{ml}^3} - \frac{i \psi^{(0)}(z_{m-}) (-2 i \Gamma + \epsilon_{nm})}{2 \pi \epsilon_{nm} \epsilon_{ml}^3} \\
		& - \frac{\psi^{(0)}(z_{l-}) (2 \Gamma + i \epsilon_{nm}) (\epsilon_{nm}^2 + 3 \epsilon_{nl} \epsilon_{ml})}{2 \pi \epsilon_{nl}^2 \epsilon_{ml}^2} + \frac{\psi^{(0)}(z_{l+}) (2 \Gamma + i \epsilon_{nm}) (\epsilon_{nm}^2 + 3 \epsilon_{nl} \epsilon_{ml})}{2 \pi \epsilon_{nl}^2 \epsilon_{ml}^2}
	\end{aligned}
\end{equation}

\begin{equation}
	\begin{aligned}
		R_1 = & -\beta \Gamma \psi^{(1)}(z_{m-}) \frac{(\Gamma - i \epsilon_{ml})}{\pi^2 \epsilon_{ml}^2 (2 i \Gamma + \epsilon_{ml})^2} + \beta \Gamma \psi^{(1)}(z_{n+}) \frac{\Gamma + i \epsilon_{nl}}{\pi^2 \epsilon_{nl}^2 (2 i \Gamma - \epsilon_{nl})^2} \\
		& + \frac{i \beta \Gamma \psi^{(1)}(z_{l+}) \left( 4 \Gamma^2 (\epsilon_{nl} + \epsilon_{ml}) + \epsilon_{nm} (2 \epsilon_{nl} + \epsilon_{ml}) \epsilon_{ml} + 2 i \Gamma (\epsilon_{nm}^2 - 3 \epsilon_{ml}^2) \right)}{2 \pi^2 \epsilon_{nl}^2 \epsilon_{ml}^2 (2 i \Gamma + \epsilon_{ml})^2} \\
		& + \frac{i \beta \Gamma \psi^{(1)}(z_{l-}) \left( 4 \Gamma^2 (\epsilon_{nl} + \epsilon_{ml}) - \epsilon_{nm} (\epsilon_{nl} + 2 \epsilon_{ml}) \epsilon_{nl} + 2 i \Gamma (-\epsilon_{nm}^2 + 3 \epsilon_{nl}^2) \right)}{2 \pi^2 \epsilon_{nl}^2 \epsilon_{ml}^2 (2 i \Gamma - \epsilon_{nl})^2}
	\end{aligned}
\end{equation}

\begin{equation}
	\begin{aligned}
		R_2 = & -\frac{\beta^2 \Gamma \psi^{(2)}(z_{m-})}{8 \pi^3 \epsilon_{ml} (2 i \Gamma + \epsilon_{ml})} + \frac{\beta^2 \Gamma \psi^{(2)}(z_{n+})}{8 \pi^3 \epsilon_{ln} (2 i \Gamma - \epsilon_{ln})} 
		+ \frac{\beta^2 \Gamma \psi^{(2)}(z_{l-}) (-2 i \Gamma + \epsilon_{nm})}{8 \pi^3 \epsilon_{nl} (-2 i \Gamma + \epsilon_{nl}) \epsilon_{lm}} + \frac{\beta^2 \Gamma \psi^{(2)}(z_{l+}) (2 i \Gamma - \epsilon_{nm})}{8 \pi^3 \epsilon_{nl} \epsilon_{lm} (-2 i \Gamma - \epsilon_{ml})}
	\end{aligned}
\end{equation}

The Taylor expansion of the coefficient with respect to $\Gamma$ is given by:
\begin{equation}
	\begin{aligned}
		{\cal C}_{nml}^{(2, 2)} &= \frac{i}{2} \Bigg( \psi^{(0)}(z_{n+}^0) \frac{1}{\pi \epsilon_{nm} \epsilon_{nl}^3} - \psi^{(0)}(z_{n-}^0) \frac{1}{\pi \epsilon_{nm} \epsilon_{nl}^3}
		+ \psi^{(0)}(z_{m-}^0) \frac{1}{\pi \epsilon_{nm} \epsilon_{ml}^3} - \psi^{(0)}(z_{m+}^0) \frac{1}{\pi \epsilon_{nm} \epsilon_{ml}^3} \\
		& + \psi^{(0)}(z_{l-}^0) \frac{(\epsilon_{nm}^2 + 3 \epsilon_{ml}^2)}{\pi \epsilon_{nl}^3 \epsilon_{lm}^3} - \psi^{(0)}(z_{l+}^0) \frac{(\epsilon_{nm}^2 + 3 \epsilon_{ml}^2)}{\pi \epsilon_{nl}^3 \epsilon_{lm}^3} \Bigg) + \mathcal{O}(\Gamma)
	\end{aligned}
\end{equation}
Notably, the fully interband coefficient contributes exclusively to the intrinsic conductivity at $\mathcal{O}(\Gamma^0)$, representing a contribution that scales with the Fermi sea ($f_n$). This result is consistent with the multiband derivations provided in Supplementary Section \ref{S-E}.

\section{Expansion of DC Conductivity in \texorpdfstring{$\Gamma$}{Gamma}}
\label{S-D}

We now compute the DC conductivity $\sigma_{abc} \equiv \lim_{\omega \to 0} \sigma_{abc}(\omega, -\omega)$ for a generic two band model using expressions from Supplementary sections~\ref{S-AA}, \ref{S-B}, and \ref{S-C}. We perform a Taylor expansion of the full DC conductivity $\sigma_{abc}$ in the relaxation rate $\Gamma$ around $\Gamma=0$:
\begin{align}
\begin{aligned}
\sigma_{abc} = \frac{1}{\Gamma^2}\sigma_{abc}^{(-2)} + \frac{1}{\Gamma}\sigma_{abc}^{(-1)} + \sigma_{abc}^{(0)} + {\cal O}(\Gamma)
\end{aligned}
\end{align}
The coefficients $\sigma_{abc}^{(k)}$ are then independent of $\Gamma$. We analyze each of these coefficients for a two-band model ($n, m \in \{1, 2\}$). For conciseness we omit summation symbols for ${\bf k}$.

\subsection{Notations and Key Identities for a Two-Band Model}
\label{S-D-1}

To make the derivations self-contained, we first establish our notation and key identities, which strictly follow those in the provided analysis note.

\subsubsection*{Definitions and Notation}
\begin{enumerate}
    \item Band Energies: $\epsilon_n$. Energy difference: $\epsilon_{nm} = \epsilon_n - \epsilon_m$. We use $\bar{n}$ to denote the band other than $n$ (e.g., $\epsilon_{n\bar{n}} = \epsilon_n - \epsilon_{\bar{n}}$).
    \item Velocity Matrix Elements: $v_{nm}^a = \langle u_n | \partial_{k_a} H | u_m \rangle$. The intraband velocity is $v_n^a \equiv v_{nn}^a$.
    \item Interband Berry Connection: $A_{nm}^a = i \langle u_n | \partial_{k_a} u_m \rangle$ for $n \neq m$.
    \item Distribution Function: Here we use the convention $f(\epsilon) = \tanh [(\beta/2)(\mu - \epsilon)]$, which comes from a complex conjugate pair of Polygamma functions in the $\Gamma\to 0$ limit. Its $k$-th derivative with respect to energy is $f_n^{(k)} \equiv d^k f_n/d\epsilon^k |_{\epsilon=\epsilon_n}$. This is related to the standard Fermi-Dirac distribution $f_0 (\epsilon)$ by $f (\epsilon) = 2 f_0 (\epsilon) - 1$, and for derivatives $k \ge 1$, $f_n^{(k)} = 2 f_{0,n}^{(k)}$. The final results in the main text are presented using $f_0$.
    \item Quantum Metric: Here we use the convention $g_{ab} = \text{Re}(A_{12}^a A_{21}^b)$.
\end{enumerate}

\subsubsection*{Key Identities}
Our derivation relies on the following standard two-band model identities:
\begin{enumerate}
    \item Feynman-Hellmann Identity: For $n \neq m$, $v_{nm}^a = -i \epsilon_{mn} A_{nm}^a$.
    \item Metric-Velocity Relation: $v_{12}^a v_{21}^b + v_{12}^b v_{21}^a = \epsilon_{12}^2 (A_{12}^a A_{21}^b + A_{12}^b A_{21}^a) = 2\epsilon_{12}^2 g_{ab}$.
    \item Diagonal Second Derivative: $v_{nn}^{ab} \equiv \langle u_n | \partial_{k_a}\partial_{k_b} H | u_n \rangle = \partial_a \partial_b \epsilon_n - 2\epsilon_{n\bar{n}} g_{ab}$.
    \item Anti-symmetric Velocity Product: $V_{ab} \equiv v_{21}^a v_{12}^b - v_{12}^a v_{21}^b = i \epsilon_{12}^2 \Omega_{ab}^1$, where $\Omega_{ab}^1 = i (A_{12}^a A_{21}^b - A_{12}^b A_{21}^a)$ is the Berry curvature of band 1.
\end{enumerate}

\subsection{The \texorpdfstring{${\cal O}(\Gamma^{-2})$}{O(Gamma-2)} (Nonlinear Drude) Contribution}

The $\sigma_{abc}^{(-2)}$ is proportional to $1/\Gamma^2$ and is
\begin{align}
\begin{aligned}
\sigma_{abc}^{(-2)} = \frac{1}{8\epsilon_{12}} [ -K_1 f_1' + K_2 f_2' ]
\end{aligned}
\end{align}
The coefficients $K_1$ and $K_2$ are composed of two parts, involving interband velocities ($K_{nA}$) and intraband second derivatives ($K_{nB}$).
\begin{align}
\begin{aligned}
K_1 &= \underbrace{v_{21}^a(v_{12}^b v_1^c + v_1^b v_{12}^c) + v_{12}^a(v_{21}^b v_1^c + v_1^b v_{21}^c)}_{K_{1A}} + \underbrace{(v_{11}^{ac} v_1^b + v_{11}^{ab} v_1^c)\epsilon_{12}}_{K_{1B}} \\
K_2 &= \underbrace{v_{21}^a(v_2^b v_{12}^c + v_{12}^b v_2^c) + v_{12}^a(v_2^b v_{21}^c + v_{21}^b v_2^c)}_{K_{2A}} \underbrace{- (v_{22}^{ac} v_2^b + v_{22}^{ab} v_2^c)\epsilon_{12}}_{K_{2B}}
\end{aligned}
\end{align}
We analyze the contributions from $K_A$ and $K_B$ separately.

\subsubsection*{Analysis of \texorpdfstring{$K_A$}{KA}}
We reorganize $K_{1A}$ and apply the Metric-Velocity Relation ($v_{12}^a v_{21}^b + v_{12}^b v_{21}^a = 2\epsilon_{12}^2 g_{ab}$).
\begin{align}
\begin{aligned}
K_{1A} &= v_1^c(v_{21}^a v_{12}^b + v_{12}^a v_{21}^b) + v_1^b(v_{21}^a v_{12}^c + v_{12}^a v_{21}^c) = 2\epsilon_{12}^2 (v_1^c g_{ab} + v_1^b g_{ac})
\end{aligned}
\end{align}
Similarly, $K_{2A} = 2\epsilon_{12}^2 (v_2^c g_{ab} + v_2^b g_{ac})$.
The contribution to $\sigma_{abc}^{(-2)}$ from these terms is $T_A^{(-2)}$:
\begin{align}
\begin{aligned}
T_A^{(-2)} &= \frac{2\epsilon_{12}^2}{8\epsilon_{12}} [ -(v_1^c g_{ab} + v_1^b g_{ac}) f_1' + (v_2^c g_{ab} + v_2^b g_{ac}) f_2' ] = \frac{\epsilon_{12}}{4} [ (g_{ab} v_2^c + g_{ac} v_2^b) f_2' - (g_{ab} v_1^c + g_{ac} v_1^b) f_1' ]
\label{TA-2}
\end{aligned}
\end{align}

\subsubsection*{Analysis of \texorpdfstring{$K_B$}{KB}}
The contribution to $\sigma_{abc}^{(-2)}$ from the $K_B$ terms is $T_B^{(-2)}$:
\begin{align}
\begin{aligned}
T_B^{(-2)} &= \frac{1}{8\epsilon_{12}} [ -K_{1B} f_1' + K_{2B} f_2' ] = \frac{\epsilon_{12}}{8\epsilon_{12}} [ -(v_{11}^{ac} v_1^b + v_{11}^{ab} v_1^c) f_1' - (v_{22}^{ac} v_2^b + v_{22}^{ab} v_2^c) f_2' ] \\
&= -\frac{1}{8} \sum_n (v_{nn}^{ac} v_n^b + v_{nn}^{ab} v_n^c) f_n'
\end{aligned}
\end{align}
We use the identity for the diagonal second derivative: $v_{nn}^{ab} = \partial_a \partial_b \epsilon_n - 2 \epsilon_{n\bar{n}} g_{ab}$.
\begin{align}
\begin{aligned}
T_B^{(-2)} = -\frac{1}{8} \sum_n \left[ (\partial_a \partial_c \epsilon_n - 2 \epsilon_{n\bar{n}} g_{ac}) v_n^b + (\partial_a \partial_b \epsilon_n - 2 \epsilon_{n\bar{n}} g_{ab}) v_n^c \right] f_n'
\end{aligned}
\end{align}
We separate this into an intraband part ($T_{B, \text{intra}}^{(-2)}$) and a geometric part ($T_{B, \text{geo}}^{(-2)}$):
\begin{align}
\begin{aligned}
T_{B}^{(-2)} &= T_{B, \text{intra}}^{(-2)} + T_{B, \text{geo}}^{(-2)},
\end{aligned}
\end{align}
with
\begin{align}
\begin{aligned}
T_{B, \text{intra}}^{(-2)} &= -\frac{1}{8} \sum_n (\partial_a \partial_c \epsilon_n v_n^b + \partial_a \partial_b \epsilon_n v_n^c) f_n' = -\frac{1}{8} \sum_n \partial_a(v_n^b v_n^c) f_n'
\end{aligned}
\end{align}
and
\begin{align}
\begin{aligned}
T_{B, \text{geo}}^{(-2)} &= -\frac{1}{8} \sum_n (-2 \epsilon_{n\bar{n}}) (g_{ac} v_n^b + g_{ab} v_n^c) f_n' = \frac{1}{4} \left[ \epsilon_{12} (g_{ac} v_1^b + g_{ab} v_1^c) f_1' + \epsilon_{21} (g_{ac} v_2^b + g_{ab} v_2^c) f_2' \right]
\end{aligned}
\end{align}
Using $\epsilon_{21} = -\epsilon_{12}$:
\begin{align}
\begin{aligned}
T_{B, \text{geo}}^{(-2)} = \frac{\epsilon_{12}}{4} \left[ (g_{ac} v_1^b + g_{ab} v_1^c) f_1' - (g_{ac} v_2^b + g_{ab} v_2^c) f_2' \right]
\label{TB-2geo}
\end{aligned}
\end{align}

\subsubsection*{Total Expression for \texorpdfstring{$\sigma_{abc}^{(-2)}$}{sigma-abc-2}}
We combine all contributions: $\sigma_{abc}^{(-2)} = T_A^{(-2)} + T_{B, \text{intra}}^{(-2)} + T_{B, \text{geo}}^{(-2)}$.
Comparing the geometric terms $T_A^{(-2)}$ [Eq.~\eqref{TA-2} and $T_{B, \text{geo}}^{(-2)}$ [Eq.~\eqref{TB-2geo}], we find they are exactly opposite:
\begin{align}
\begin{aligned}
T_A^{(-2)} + T_{B, \text{geo}}^{(-2)} = 0
\end{aligned}
\end{align}
The total conductivity $\sigma_{abc}^{(-2)}$ is therefore purely intraband:
\begin{align}
\begin{aligned}
\sigma_{abc}^{(-2)} = T_{B, \text{intra}}^{(-2)} = -\frac{1}{8} \sum_{n=1,2} \partial_a(v_n^b v_n^c) f_n',
\qquad
f_n' = 2 f_{0,n}'
\end{aligned}
\end{align}
This result demonstrates that the leading divergence ($1/\Gamma^2$) depends only on the band structure derivatives (generalized Drude weight) and is independent of geometric quantities.

\subsection{The \texorpdfstring{${\cal O}(\Gamma^{-1})$}{O(Gamma-1)} (Berry Curvature Dipole) Contribution}

The term proportional to $1/\Gamma$ is proportional to the first derivative of the distribution function, $f'$.

\begin{align}
\begin{aligned}
\sigma_{abc}^{(-1)} = -\frac{i}{4\epsilon_{12}^2} [ C_1 f_1' + C_2 f_2' ]
\label{sigma_abc_-1}
\end{aligned}
\end{align}
The coefficients $C_1$ and $C_2$ are given by:
\begin{align}
\begin{aligned}
C_1 &= v_{21}^a(v_{12}^b v_1^c + v_1^b v_{12}^c) - v_{12}^a(v_{21}^b v_1^c + v_1^b v_{21}^c) \\
C_2 &= -v_{21}^a(v_2^b v_{12}^c + v_{12}^b v_2^c) + v_{12}^a(v_2^b v_{21}^c + v_{21}^b v_2^c)
\end{aligned}
\end{align}
We reorganize these expressions by factoring out the intraband velocities:
\begin{align}
\begin{aligned}
C_1 &= v_1^c (v_{21}^a v_{12}^b - v_{12}^a v_{21}^b) + v_1^b (v_{21}^a v_{12}^c - v_{12}^a v_{21}^c) \\
C_2 &= v_2^c (v_{12}^a v_{21}^b - v_{21}^a v_{12}^b) + v_2^b (v_{12}^a v_{21}^c - v_{21}^a v_{12}^c)
\label{C1C2-1}
\end{aligned}
\end{align}

The anti-symmetric structure suggests a connection to the Berry curvature. We analyze the velocity combination $V_{ab} \equiv v_{21}^a v_{12}^b - v_{12}^a v_{21}^b$. We utilize the Feynman-Hellmann identity ($v_{nm}^a = -i \epsilon_{mn} A_{nm}^a$).
\begin{align}
\begin{aligned}
V_{ab} &= (-i\epsilon_{12} A_{21}^a)(-i\epsilon_{21} A_{12}^b) - (-i\epsilon_{21} A_{12}^a)(-i\epsilon_{12} A_{21}^b) = (-1)(\epsilon_{12}\epsilon_{21}) [ A_{21}^a A_{12}^b - A_{12}^a A_{21}^b ] \\
&= \epsilon_{12}^2 (A_{21}^a A_{12}^b - A_{12}^a A_{21}^b) 
\qquad \qquad
(\epsilon_{12}\epsilon_{21} = -\epsilon_{12}^2) .
\label{Vab-1}
\end{aligned}
\end{align}

The Berry curvature for band $n$ in a two-band system is defined as $\Omega_{ab}^n = i (A_{n\bar{n}}^a A_{\bar{n}n}^b - A_{n\bar{n}}^b A_{\bar{n}n}^a)$.
For band 1: $\Omega_{ab}^1 = i (A_{12}^a A_{21}^b - A_{12}^b A_{21}^a)$.
We relate the term in $V_{ab}$ to $\Omega_{ab}^1$:
\begin{align}
\begin{aligned}
A_{21}^a A_{12}^b - A_{12}^a A_{21}^b = -(A_{12}^a A_{21}^b - A_{12}^b A_{21}^a) = -(-i \Omega_{ab}^1) = i \Omega_{ab}^1
\end{aligned}
\end{align}
Substituting this back into $V_{ab}$ [Eq.~\eqref{Vab-1}]:
\begin{align}
\begin{aligned}
V_{ab} = i \epsilon_{12}^2 \Omega_{ab}^1
\end{aligned}
\end{align}

We substitute the expression for $V_{ab}$ back into the coefficients $C_1$ and $C_2$ [Eq.~\eqref{C1C2-1}].
\begin{align}
\begin{aligned}
C_1 = i \epsilon_{12}^2 (v_1^c \Omega_{ab}^1 + v_1^b \Omega_{ac}^1),
\qquad
C_2 = -i \epsilon_{12}^2 (v_2^c \Omega_{ab}^1 + v_2^b \Omega_{ac}^1)
\end{aligned}
\end{align}
Finally, we substitute $C_1$ and $C_2$ into the expression for $\sigma_{abc}^{(-1)}$ [Eq.~\eqref{sigma_abc_-1}]:
\begin{align}
\begin{aligned}
\sigma_{abc}^{(-1)} &= -\frac{i}{4\epsilon_{12}^2} (i \epsilon_{12}^2) \left[ (v_1^c \Omega_{ab}^1 + v_1^b \Omega_{ac}^1) f_1' - (v_2^c \Omega_{ab}^1 + v_2^b \Omega_{ac}^1) f_2' \right] \\
&= \frac{1}{4} \left[ (v_1^c \Omega_{ab}^1 + v_1^b \Omega_{ac}^1) f_1' - (v_2^c \Omega_{ab}^1 + v_2^b \Omega_{ac}^1) f_2' \right]
\end{aligned}
\end{align}
Using the property that the Berry curvature is opposite for the two bands ($\Omega_{ab}^2 = -\Omega_{ab}^1$), we can write this compactly:
\begin{align}
\begin{aligned}
\sigma_{abc}^{(-1)} = \frac{1}{4} \sum_{n=1,2} (v_n^c \Omega_{ab}^n + v_n^b \Omega_{ac}^n) f_n',
\qquad
f_n' = 2f_{0,n}'
\end{aligned}
\end{align}
This term is a Fermi surface contribution related to the nonlinear anomalous Hall effect (NAHE).

\subsection{The \texorpdfstring{${\cal O}(\Gamma^{0})$}{O(Gamma-0)} (Intrinsic) Contribution}
\label{Two-band Intrinsic}

This is the central result. The $\Gamma^0$ term, $\sigma_{abc}^{(0)}$, is composed of five distinct terms from the $\Gamma$-expansion, which we group based on the power of the inverse temperature $\beta$:
$\sigma_{abc}^{(0)} = T_0 + T_1 + T_2 + T_3 + T_4$.

\subsubsection*{Term 4 (\texorpdfstring{$\sim \beta^4$}{~beta4}, \texorpdfstring{$f^{(4)}$}{f4})}
This term is proportional to the fourth derivative of $f(\epsilon)$:
\begin{align}
\begin{aligned}
T_4 = \sum_{n=1,2} \frac{1}{4!} v_n^a v_n^b v_n^c f_n^{(4)} = \sum_{n=1,2} \frac{1}{24} v_n^a v_n^b v_n^c f_n^{(4)},
\end{aligned}
\end{align}
which is a purely intraband contribution.

\subsubsection*{Term 3 (\texorpdfstring{$\sim \beta^3$}{beta3}, \texorpdfstring{$f^{(3)}$}{f3})}
This term is proportional to the third derivative.
\begin{align}
\begin{aligned}
T_3 = \frac{1}{48\epsilon_{12}} \left[ K_1 f_1^{(3)} - K_2 f_2^{(3)} \right]
\end{aligned}
\end{align}
The coefficients $K_n$ contain terms involving interband velocities ($K_{nA}$) and second derivative matrix elements ($K_{nB}$).
\begin{align}
\begin{aligned}
K_1 &= \underbrace{v_{21}^a(v_{12}^b v_1^c + v_1^b v_{12}^c) + v_{12}^a(v_{21}^b v_1^c + v_1^b v_{21}^c)}_{K_{1A}} + \underbrace{(v_{11}^{ac}v_1^b + v_{11}^{ab}v_1^c)\epsilon_{12}}_{K_{1B}} \\
K_2 &= \underbrace{v_{21}^a(v_2^b v_{12}^c + v_{12}^b v_2^c) + v_{12}^a(v_2^b v_{21}^c + v_{21}^b v_2^c)}_{K_{2A}} \underbrace{- (v_{22}^{ac}v_2^b + v_{22}^{ab}v_2^c)\epsilon_{12}}_{K_{2B}}
\end{aligned}
\end{align}
We analyze the contributions $T_{3A}$ and $T_{3B}$ separately. Using the Metric-Velocity Relation, $K_{1A} = 2\epsilon_{12}^2 (v_1^c g_{ab} + v_1^b g_{ac})$ and $K_{2A} = 2\epsilon_{12}^2 (v_2^c g_{ab} + v_2^b g_{ac})$.

For $T_{3A}$,
\begin{align}
\begin{aligned}
T_{3A} &= \frac{2\epsilon_{12}^2}{48\epsilon_{12}} \left[ (v_1^c g_{ab} + v_1^b g_{ac}) f_1^{(3)} - (v_2^c g_{ab} + v_2^b g_{ac}) f_2^{(3)} \right] = \frac{\epsilon_{12}}{24} \left[ (g_{ab} v_1^c + g_{ac} v_1^b) f_1^{(3)} - (g_{ab} v_2^c + g_{ac} v_2^b) f_2^{(3)} \right]
\end{aligned}
\end{align}

For $T_{3B}$,
\begin{align}
\begin{aligned}
T_{3B} &= \frac{1}{48\epsilon_{12}} \left[ K_{1B} f_1^{(3)} - K_{2B} f_2^{(3)} \right] = \frac{\epsilon_{12}}{48\epsilon_{12}} \left[ (v_{11}^{ac}v_1^b + v_{11}^{ab}v_1^c) f_1^{(3)} + (v_{22}^{ac}v_2^b + v_{22}^{ab}v_2^c) f_2^{(3)} \right] \\
&= \frac{1}{48} \sum_n (v_{nn}^{ac}v_n^b + v_{nn}^{ab}v_n^c) f_n^{(3)}
\end{aligned}
\end{align}

We utilize the Diagonal Second Derivative Identity: $v_{nn}^{ab} = \partial_a \partial_b \epsilon_n - 2 \epsilon_{n\bar{n}} g_{ab}$.
\begin{align}
\begin{aligned}
T_{3B} = \frac{1}{48} \sum_n & \left[ (\partial_a \partial_c \epsilon_n - 2 \epsilon_{n\bar{n}} g_{ac})v_n^b + (\partial_a \partial_b \epsilon_n - 2 \epsilon_{n\bar{n}} g_{ab})v_n^c \right] f_n^{(3)} ,
\end{aligned}
\end{align}
and separate this into intraband derivative terms ($T_{3B}^{\text{intra}}$) and geometric terms ($T_{3B}^{\text{inter}}$):
\begin{align}
\begin{aligned}
T_{3B}^{\text{intra}} &= \frac{1}{48} \sum_n (\partial_a \partial_c \epsilon_n v_n^b + \partial_a \partial_b \epsilon_n v_n^c) f_n^{(3)}
\end{aligned}
\end{align}
\begin{align}
\begin{aligned}
T_{3B}^{\text{inter}} &= -\frac{2}{48} \sum_n \epsilon_{n\bar{n}} (g_{ac}v_n^b + g_{ab}v_n^c) f_n^{(3)}
\end{aligned}
\end{align}
Since $\epsilon_{21} = -\epsilon_{12}$, we find $T_{3B}^{\text{inter}} = -T_{3A}$.

For Total $T_3$, the geometric contributions cancel exactly: $T_3 = T_{3A} + T_{3B} = T_{3B}^{\text{intra}}$. We recognize the intraband term as a total derivative: $\partial_a \partial_c \epsilon_n v_n^b + \partial_a \partial_b \epsilon_n v_n^c = \partial_a(v_n^b v_n^c)$.
\begin{align}
\begin{aligned}
T_3 = \frac{1}{48} \sum_{n=1,2} \partial_a(v_n^b v_n^c) f_n^{(3)}
\label{Term3-0}
\end{aligned}
\end{align}
This is also a purely intraband contribution.

\subsubsection*{Term 2 (\texorpdfstring{$\sim \beta^2$}{beta2}, \texorpdfstring{$f^{(2)}$}{f2})}
This term is proportional to the second derivative.
\begin{align}
\begin{aligned}
T_2 = \frac{1}{4\epsilon_{12}^2} \sum_n v_n^a (v_{21}^b v_{12}^c + v_{12}^b v_{21}^c) f_n^{(2)}
\end{aligned}
\end{align}
Applying the Metric-Velocity Relation:
\begin{align}
\begin{aligned}
T_2 &= \frac{1}{4\epsilon_{12}^2} \sum_n v_n^a (2\epsilon_{12}^2 g_{bc}) f_n^{(2)} = \frac{1}{2} g_{bc} \sum_n v_n^a f_n^{(2)}
\label{Term2-0}
\end{aligned}
\end{align}

\subsubsection*{Term 1 (\texorpdfstring{$\sim \beta^1$}{beta1}, \texorpdfstring{$f^{(1)}$}{f1})}
This term is proportional to the first derivative. By matching coefficients:
\begin{align}
\begin{aligned}
T_1 = \frac{1}{2\epsilon_{12}^3} \left[ C_1 f_1^{(1)} - C_2 f_2^{(1)} \right]
\end{aligned}
\end{align}
The coefficients $C_1$ and $C_2$ are:
\begin{align}
\begin{aligned}
C_1 &= v_{21}^a(v_{12}^b v_1^c + v_1^b v_{12}^c) + v_{12}^a(v_{21}^b v_1^c + v_1^b v_{21}^c) - v_1^a(v_{21}^b v_{12}^c + v_{12}^b v_{21}^c) \\
C_2 &= v_{21}^a(v_2^b v_{12}^c + v_{12}^b v_2^c) + v_{12}^a(v_2^b v_{21}^c + v_{21}^b v_2^c) - v_2^a(v_{21}^b v_{12}^c + v_{12}^b v_{21}^c)
\end{aligned}
\end{align}
We reorganize $C_1$ and apply the Metric-Velocity Relation:
\begin{align}
\begin{aligned}
C_1 &= v_1^c(v_{21}^a v_{12}^b + v_{12}^a v_{21}^b) + v_1^b(v_{21}^a v_{12}^c + v_{12}^a v_{21}^c) - v_1^a(v_{21}^b v_{12}^c + v_{12}^b v_{21}^c) \\
&= v_1^c(2\epsilon_{12}^2 g_{ab}) + v_1^b(2\epsilon_{12}^2 g_{ac}) - v_1^a(2\epsilon_{12}^2 g_{bc})
\end{aligned}
\end{align}
$C_2$ follows similarly. Substituting these back into $T_1$:
\begin{align}
\begin{aligned}
T_1 &= \frac{2\epsilon_{12}^2}{2\epsilon_{12}^3} \left[ (v_1^c g_{ab} + v_1^b g_{ac} - v_1^a g_{bc}) f_1' - (v_2^c g_{ab} + v_2^b g_{ac} - v_2^a g_{bc}) f_2' \right] \\
&= \frac{1}{\epsilon_{12}} \left[ (v_1^c g_{ab} + v_1^b g_{ac} - v_1^a g_{bc}) f_1' - (v_2^c g_{ab} + v_2^b g_{ac} - v_2^a g_{bc}) f_2' \right]
\label{Term1-0}
\end{aligned}
\end{align}

\subsubsection*{Term 0 (\texorpdfstring{$\sim \beta^0$}{beta0}, \texorpdfstring{$f$}{f0})}
This term represents a Fermi sea contribution at first glance:
\begin{align}
\begin{aligned}
T_0 = \frac{f_1 - f_2}{2\epsilon_{12}^4} \times C_0
\label{Term0}
\end{aligned}
\end{align}
The coefficient $C_0$ is decomposed into three parts ($A$, $B$, $E$):

\begin{align}
\begin{aligned}
A &= 3 (v_2^a - v_1^a) (v_{21}^b v_{12}^c + v_{12}^b v_{21}^c) \\
B &= (v_2^b-v_1^b)(v_{21}^a v_{12}^c + v_{12}^a v_{21}^c) + (v_2^c-v_1^c)(v_{21}^a v_{12}^b + v_{12}^a v_{21}^b) \\
E &= (v_{21}^{ac}v_{12}^b + v_{12}^{ac}v_{21}^b + v_{21}^{ab}v_{12}^c + v_{12}^{ab}v_{21}^c)\epsilon_{12}
\end{aligned}
\end{align}

For Part A and Part B we use Metric-Velocity Relation
\begin{align}
\begin{aligned}
A = 3 (v_2^a - v_1^a) (2\epsilon_{12}^2 g_{bc}) = 6\epsilon_{12}^2 g_{bc} (v_2^a - v_1^a)
\label{A_Term0}
\end{aligned}
\end{align}
\begin{align}
\begin{aligned}
B = 2\epsilon_{12}^2 [ g_{ac}(v_2^b-v_1^b) + g_{ab}(v_2^c-v_1^c) ]
\label{B_Term0}
\end{aligned}
\end{align}

For Part E, we define $S_{ab;c} = v_{21}^{ab}v_{12}^c + v_{12}^{ab}v_{21}^c$. Then $E = (S_{ac;b} + S_{ab;c}) \epsilon_{12}$.
We use the Off-Diagonal Second Derivative Identity: $v_{nm}^{ab} = \partial_a v_{nm}^b - i [A^a, v^b]_{nm}$.
\begin{align}
\begin{aligned}
S_{ab;c} &= (\partial_a v_{21}^b) v_{12}^c + (\partial_a v_{12}^b) v_{21}^c - i K_{ab;c} = \partial_a(v_{21}^b v_{12}^c) - i K_{ab;c}
\end{aligned}
\end{align}
where $K_{ab;c} = [A^a, v^b]_{21} v_{12}^c + [A^a, v^b]_{12} v_{21}^c$.
We consider the symmetric combination required for E:
\begin{align}
\begin{aligned}
S_{ac;b} + S_{ab;c} &= \partial_a(v_{21}^c v_{12}^b + v_{21}^b v_{12}^c) - i K_{abc}
\end{aligned}
\end{align}
where $K_{abc} = K_{ac;b} + K_{ab;c}$. We use Metric-Velocity Relation for the term inside the derivative:
\begin{align}
\begin{aligned}
S_{ac;b} + S_{ab;c} = \partial_a (2 \epsilon_{12}^2 g_{bc}) - i K_{abc}
\end{aligned}
\end{align}

We evaluate the commutator term $K_{abc}$ and check its gauge invariance. In a two-band system, the commutators are:
\begin{align}
\begin{aligned}
[A^a, v^b]_{12} &= A_{12}^a (v_2^b - v_1^b) + v_{12}^b (A_{11}^a - A_{22}^a) \\
[A^a, v^b]_{21} &= A_{21}^a (v_1^b - v_2^b) + v_{21}^b (A_{22}^a - A_{11}^a)
\end{aligned}
\end{align}
Substituting these into the definition of $K_{ab;c}$:
\begin{align}
\begin{aligned}
K_{ab;c} = (v_2^b - v_1^b) (A_{12}^a v_{21}^c - A_{21}^a v_{12}^c) + (A_{11}^a - A_{22}^a) (v_{12}^b v_{21}^c - v_{21}^b v_{12}^c)
\end{aligned}
\end{align}
When calculating $K_{abc} = K_{ac;b} + K_{ab;c}$, the gauge-dependent part (proportional to $A_{11}^a - A_{22}^a$) is:
\begin{align}
\begin{aligned}
P_\text{gauge} = (A_{11}^a - A_{22}^a) [ (v_{12}^c v_{21}^b - v_{21}^c v_{12}^b) + (v_{12}^b v_{21}^c - v_{21}^b v_{12}^c) ]
\end{aligned}
\end{align}
Since the velocity matrix elements commute ($v_{nm}^a v_{kl}^b = v_{kl}^b v_{nm}^a$), the terms inside the bracket cancel exactly: $P_\text{gauge}=0$. This explicitly confirms the gauge invariance of E.

The remaining gauge-invariant part of $K_{abc}$ is:
\begin{align}
\begin{aligned}
K_{abc} = (v_2^c - v_1^c) (A_{12}^a v_{21}^b - A_{21}^a v_{12}^b) + (v_2^b - v_1^b) (A_{12}^a v_{21}^c - A_{21}^a v_{12}^c)
\end{aligned}
\end{align}
We evaluate the bracketed terms using the Feynman-Hellmann identity.
\begin{align}
\begin{aligned}
A_{12}^a v_{21}^b - A_{21}^a v_{12}^b &= A_{12}^a (-i\epsilon_{12} A_{21}^b) - A_{21}^a (-i\epsilon_{21} A_{12}^b) = -i\epsilon_{12} (A_{12}^a A_{21}^b + A_{21}^a A_{12}^b) = -2i\epsilon_{12} g_{ab}
\end{aligned}
\end{align}
Substituting this back into $K_{abc}$:
\begin{align}
\begin{aligned}
K_{abc} = -2i \epsilon_{12} \left[ (v_2^c - v_1^c) g_{ab} + (v_2^b - v_1^b) g_{ac} \right]
\end{aligned}
\end{align}
Now we assemble the expression for $S_{ac;b} + S_{ab;c}$ (Eq. 28):
\begin{align}
\begin{aligned}
S_{ac;b} + S_{ab;c} &= \partial_a (2 \epsilon_{12}^2 g_{bc}) - i K_{abc} = \partial_a (2 \epsilon_{12}^2 g_{bc}) - 2 \epsilon_{12} \left[ g_{ab} (v_2^c - v_1^c) + g_{ac} (v_2^b - v_1^b) \right]
\end{aligned}
\end{align}
We expand the derivative term, using $\partial_a \epsilon_{12} = v_1^a - v_2^a$:
\begin{align}
\begin{aligned}
\partial_a (2 \epsilon_{12}^2 g_{bc}) &= 4 \epsilon_{12} (\partial_a \epsilon_{12}) g_{bc} + 2 \epsilon_{12}^2 (\partial_a g_{bc}) = 4 \epsilon_{12} (v_1^a - v_2^a) g_{bc} + 2 \epsilon_{12}^2 (\partial_a g_{bc})
\end{aligned}
\end{align}
The complete expression for $E = (S_{ac;b} + S_{ab;c}) \epsilon_{12}$ is:
\begin{align}
\begin{aligned}
E &= 4 \epsilon_{12}^2 (v_1^a - v_2^a) g_{bc} + 2 \epsilon_{12}^3 (\partial_a g_{bc}) - 2 \epsilon_{12}^2 \left[ g_{ab} (v_2^c - v_1^c) + g_{ac} (v_2^b - v_1^b) \right]
\label{E_Term0}
\end{aligned}
\end{align}

We sum the results A [Eq.~\eqref{A_Term0}], B [Eq.~\eqref{B_Term0}], and E [Eq.~\eqref{E_Term0}]:
\begin{align}
\begin{aligned}
C_0 &= \underbrace{6\epsilon_{12}^2 g_{bc} (v_2^a - v_1^a)}_{A} + \underbrace{2\epsilon_{12}^2 [ g_{ac}(v_2^b-v_1^b) + g_{ab}(v_2^c-v_1^c) ]}_{B} + E
\end{aligned}
\end{align}
We group the terms by the geometric quantities:
\begin{align}
\begin{aligned}
C_0 &= \epsilon_{12}^2 g_{bc} [ 6(v_2^a-v_1^a) + 4(v_1^a-v_2^a) ] \\
& \quad + \epsilon_{12}^2 g_{ac} [ 2(v_2^b-v_1^b) - 2(v_2^b-v_1^b) ] \\
& \quad + \epsilon_{12}^2 g_{ab} [ 2(v_2^c-v_1^c) - 2(v_2^c-v_1^c) ] + 2\epsilon_{12}^3 (\partial_a g_{bc})
\end{aligned}
\end{align}
The terms proportional to $g_{ac}$ and $g_{ab}$ cancel exactly.
\begin{align}
\begin{aligned}
C_0 &= 2\epsilon_{12}^2 g_{bc} (v_2^a-v_1^a) + 2\epsilon_{12}^3 (\partial_a g_{bc})
\end{aligned}
\end{align}

Substituting the simplified $C_0$ back into $T_0$ [Eq.~\eqref{Term0}]:
\begin{align}
\begin{aligned}
T_0 &= (f_1 - f_2) \left[ \frac{g_{bc} (v_2^a-v_1^a)}{\epsilon_{12}^2} + \frac{\partial_a g_{bc}}{\epsilon_{12}} \right]
\end{aligned}
\end{align}
Since $v_2^a-v_1^a = -(\partial_a \epsilon_{12})$, this expression is recognized as the quotient rule for a total derivative:
\begin{align}
\begin{aligned}
T_0 = (f_1 - f_2) \left[ \frac{(\partial_a g_{bc})\epsilon_{12} - g_{bc}(\partial_a \epsilon_{12})}{\epsilon_{12}^2} \right] = (f_1 - f_2) \partial_a \left( \frac{g_{bc}}{\epsilon_{12}} \right) .
\label{Term0-0}
\end{aligned}
\end{align}

\subsection*{III. Transformation and Final Expressions}
\label{Two-band Final Expression}

We use integration by parts (IBP) in $k$-space, $\int (\partial_a A) B = - \int A (\partial_a B)$, to simplify the expressions and group them into physical Fermi surface contributions.

\subsubsection*{A. Kinetic Contribution}
\label{A-Kinetic-Contribution}
The total kinetic contribution is $\sigma_{abc}^{\text{kin}} = T_3 + T_4$.
We apply IBP to $T_3$ [Eq.~\eqref{Term3-0}] with respect to $k_a$.
\begin{align}
\begin{aligned}
T_3 \quad \xrightarrow{\text{IBP}} \quad & -\frac{1}{48} \sum_n (v_n^b v_n^c) \partial_a (f_n^{(3)})
\end{aligned}
\end{align}
Using the chain rule $\partial_a f_n^{(3)} = f_n^{(4)} v_n^a$:
\begin{align}
\begin{aligned}
T_3 \quad \xrightarrow{\text{IBP}} \quad & -\frac{1}{48} \sum_n v_n^a v_n^b v_n^c f_n^{(4)} = -\frac{1}{2} T_4
\end{aligned}
\end{align}
The total intraband contribution simplifies to:
\begin{align}
\begin{aligned}
\sigma_{abc}^{\text{kin}} = T_3 + T_4 = \frac{1}{2}T_4 = \sum_{n=1,2} \frac{1}{48} v_n^a v_n^b v_n^c f_n^{(4)}
=\sum_{n=1,2} \frac{1}{24} v_n^a v_n^b v_n^c f_{0,n}^{(4)} ,
\quad
f_n' = 2 f_{0,n}'
\end{aligned}
\end{align}
This result highlights that the specific mechanism of dissipation (fermionic bath) dictates the NESS structure such that a purely kinetic, ballistic response survives in the intrinsic $\left(\Gamma^0\right)$ limit, originating from the Polygamma functions discussed in Section~\ref{Fully-Intraband-Coefficient}. This contrasts with models assuming RTAs.

To improve numerical stability, we can reduce the order of the derivative on the Fermi function by integration by parts over the Brillouin Zone, and rewrite $\sigma^{\text{kin}}_{abc}$ as: 
\begin{equation}
\sigma^{\text{kin}}_{abc} = \sum_{n} \frac{1}{12} \left( \frac{\partial^3 \epsilon_n}{\partial k_a \partial k_b \partial k_c} \right) f_{0,n}^{(2)}.
\end{equation}

\subsubsection*{B. Geometric Contribution}
We apply IBP to $T_0$ [Eq.~\eqref{Term0-0}] to transform the Fermi sea term to the Fermi surface:
\begin{align}
\begin{aligned}
T_0 \quad \xrightarrow{\text{IBP}} \quad & - \left[ \partial_a (f_1 - f_2) \right] \left( \frac{g_{bc}}{\epsilon_{12}} \right) = - (f_1' v_1^a - f_2' v_2^a) \left( \frac{g_{bc}}{\epsilon_{12}} \right) = g_{bc} \left( \frac{v_2^a f_2' - v_1^a f_1'}{\epsilon_{12}} \right),
\end{aligned}
\end{align}
The total interband Fermi surface contribution is $\sigma_{abc}^{\text{inter-FS}} = T_0 + T_1 + T_2$.
We combine the $f'$ terms from $T_0$ and $T_1$ [Eq.~\eqref{Term1-0}]:
\begin{align}
\begin{aligned}
T_0 + T_1 &= \frac{1}{\epsilon_{12}} \left[ g_{bc}(v_2^a f_2' - v_1^a f_1') + (v_1^c g_{ab} + v_1^b g_{ac} - v_1^a g_{bc}) f_1' - (v_2^c g_{ab} + v_2^b g_{ac} - v_2^a g_{bc}) f_2' \right]
\label{Term0+Term1-0}
\end{aligned}
\end{align}
Grouping terms by $f_1'$ and $f_2'$:
\begin{align}
\begin{aligned}
T_0 + T_1 &= \frac{1}{\epsilon_{12}} \left[ f_1' (v_1^c g_{ab} + v_1^b g_{ac} - 2v_1^a g_{bc}) - f_2' (v_2^c g_{ab} + v_2^b g_{ac} - 2v_2^a g_{bc}) \right]
\end{aligned}
\end{align}
The total geometric contribution, including $T_2$ [Eq.~\eqref{Term2-0}], is:
\begin{align}
\begin{aligned}
\sigma_{abc}^{\text{geo}} = \frac{1}{2} g_{bc} \sum_n v_n^a f_n'' + (T_0+T_1) 
\end{aligned}
\end{align}
We transform the $f''$ term ($T_2$) into an $f'$ term using IBP. We use the identity $v_n^a f_n^{(2)} = \partial_a(f_n')$.
\begin{align}
\begin{aligned}
T_2 = \sum_n (\frac{1}{2} g_{bc}) \partial_a(f_n')
\end{aligned}
\end{align}
Applying IBP w.r.t $k_a$:
\begin{align}
\begin{aligned}
T_2 \quad \xrightarrow{\text{IBP}} \quad & - \sum_n \left[ \partial_a (\frac{1}{2} g_{bc}) \right] f_n' = -\frac{1}{2} (\partial_a g_{bc}) (f_1' + f_2')
\end{aligned}
\end{align}
This transformation explicitly reveals the intraband quantum metric contribution, $-\partial_a g_{b c}/2$. Physically, this term corresponds to the Fermi surface contribution arising from the gradient of the quantum geometry, which governs the deformation and geometric shift of wave packets under the external field.

Substituting this transformed $T_2$ and $(T_0+T_1)$ [Eq.~\eqref{Term0+Term1-0}] into $\sigma_{abc}^{\text{geo}}$ and grouping terms by $f_n'$:
\begin{align}
\begin{aligned}
\sigma_{abc}^{\text{geo}} = & \quad f_1' \left[ -\frac{1}{2} (\partial_a g_{bc}) + \frac{1}{\epsilon_{12}} (v_1^c g_{ab} + v_1^b g_{ac} - 2v_1^a g_{bc}) \right] \\
& + f_2' \left[ -\frac{1}{2} (\partial_a g_{bc}) - \frac{1}{\epsilon_{12}} (v_2^c g_{ab} + v_2^b g_{ac} - 2v_2^a g_{bc}) \right]
\end{aligned}
\end{align}
We rewrite the expression for band 2 using the energy denominator $\epsilon_{21} = -\epsilon_{12}$ to achieve a symmetric form:
\begin{align}
\begin{aligned}
\sigma_{abc}^{\text{geo}} = & \quad f_1' \left[ -\frac{1}{2} (\partial_a g_{bc}) + \frac{1}{\epsilon_{12}} (g_{ab} v_1^c + g_{ac} v_1^b - 2 g_{bc} v_1^a) \right] \\
& + f_2' \left[ -\frac{1}{2} (\partial_a g_{bc}) + \frac{1}{\epsilon_{21}} (g_{ab} v_2^c + g_{ac} v_2^b - 2 g_{bc} v_2^a) \right]
\end{aligned}
\end{align}
This can be written compactly using $\epsilon_{n\bar{n}}$ (where $\epsilon_{1\bar{1}}=\epsilon_{12}$ and $\epsilon_{2\bar{2}}=\epsilon_{21}$):
\begin{align}
\label{Two-band Geometric}
\begin{aligned}
\sigma_{abc}^{\text{geo}} = \sum_{n=1,2} 
2 f_{0,n}'\left[ -\frac{1}{2} (\partial_a g_{bc}) + \frac{1}{\epsilon_{n\bar{n}}} (g_{ab} v_n^c + g_{ac} v_n^b - 2 g_{bc} v_n^a) \right],
\quad
f_n' = 2 f_{0,n}'
\end{aligned}
\end{align}

\newpage

\section{DC Conductivity of  Multiband Model}
\label{S-E}
We now consider a general multiband model to calculate the DC conductivity $\sigma_{abc} = \lim_{\omega \rightarrow 0}\sigma_{abc}(\omega, -\omega)$, and demonstrate that the results obtained for the two-band model in Supplementary Section \ref{S-D} can be generalized to an multiband system. The derivation procedure remains essentially similar to that presented in Supplementary Section \ref{S-D}. In the following derivation, we utilize abstract band indices ($n, l, m \in \{1, 2, ..., N\}$), rather than the specific indices.

\subsection{Notation and Key Identities for a Multiband Model}
In Supplementary Section \ref{S-D-1}, we defined the notations and key identities specific to the two-band model. However, to investigate multiband systems, this formalism must be further generalized. We summarize the necessary modifications and extensions below:

\subsubsection*{Definitions and Notation}
\begin{enumerate}
	\item Group Velocity Difference: $\Delta^a_{nm} = v^a_n - v^a_m$.
    \item Summation over band indices: $\sum_{I}'$, where $I$ represents the set of indices to be summed, and the prime ($^\prime$) in the superscript denotes that all indices in the expression must be mutually distinct.
	\item Multiband Berry Curvature: $\Omega^{ab}_{n} = i\sum_{m}' (A^a_{nm}A^b_{mn} - A^b_{nm}A^a_{mn})$.
	\item Multiband Quantum Metric: $g^{ab}_n = \sum_{m}'\text{Re}(A^a_{nm}A^{b}_{mn})$.
	\item Multiband Band-renormalized Quantum Metric: $\mathcal{G}^{ab}_n = \sum_{m}' \text{Re}(A^a_{nm}A^{b}_{mn}/\epsilon_{nm})$.
\end{enumerate}

\subsubsection*{Key Identities}
\begin{enumerate}
	\item Multiband Metric-Velocity Relation: $\sum_{m}'(v^{a}_{nm}v^b_{mn} +  v^b_{nm}v^a_{mn})/\epsilon^2_{nm} =  2g^{ab}_{n}$.
	\item Multiband Band-renormalized Metric-Velocity Relation: $\sum_{m}'(v^{a}_{nm}v^b_{mn} +  v^b_{nm}v^a_{mn})/\epsilon^3_{nm} =  2\mathcal{G}^{ab}_{n}$.
	\item Multiband Diagonal Second Derivative:
    \begin{equation}
		v^{ab}_{nn} = \partial_a \partial_b \epsilon_n - {\textstyle \sum_{m}'}\frac{v^a_{nm}v^b_{mn} + v^b_{nm}v^a_{mn}}{\epsilon_{nm}}.
	\end{equation}
	
	\item Multiband Off-diagonal Second Derivative:
	\begin{equation}
		v^{ab}_{nm} = -iR^{ab}_{nm}v^b_{nm} + \frac{v^b_{nm}\Delta^a_{nm}}{\epsilon_{nm}} + \frac{v^a_{nm}\Delta^b_{nm}}{\epsilon_{nm}} - 
        {\textstyle \sum_{l}'}
        \Big(\frac{v^a_{nl} v^b_{lm}}{\epsilon_{nl}} - \frac{v^b_{nl}v^a_{lm}}{\epsilon_{lm}}\Big).
	\end{equation}
	where the shift vector is defined as $R^{ab}_{nm} = i\partial_{a} \ln A^{b}_{nm} + A^{a}_{nn} - A^{a}_{mm}$.
	
	\item Multiband Curvature-Velocity Relation: $\sum_{m}'(v^{a}_{nm}v^b_{mn} -  v^b_{nm}v^a_{mn})/\epsilon^2_{nm} =  -i\Omega^{ab}_{n}$.
\end{enumerate}

\subsection{The \texorpdfstring{${\cal O}(\Gamma^{-2})$}{O(Gamma-2)} (Nonlinear Drude) Contribution}
The $\Gamma^{-2}$ contribution, $\sigma^{(-2)}_{abc}$ is given by:
\begin{equation}
	\begin{aligned}
		\sigma^{(-2)}_{abc} & =-\frac{1}{8}
        {\textstyle \sum_{n}}f^{(1)}_n \Big[ v^{ac}_{nn}v^b_{n} + v^{ab}_{nn}v^c_{n} + 
        {\textstyle \sum_{m}'}
        \Big( v^c_{n}\frac{v^{a}_{mn}v^{b}_{nm} + v^{b}_{mn}v^{a}_{nm}}{\epsilon_{nm}} + v^b_{n}\frac{v^{a}_{mn}v^{c}_{nm} + v^{c}_{mn}v^{a}_{nm}}{\epsilon_{nm}}\Big) \Big]\\ 
	\end{aligned}
\end{equation}

By using the property of Diagonal Second Derivative, the expression can be significantly simplified to:
\begin{equation}
	\begin{aligned}
		\sigma^{(-2)}_{abc}
		&= -\frac{1}{8}
        {\textstyle \sum_{n}}f^{(1)}_n \Big(\partial_a\partial_c\epsilon_n v^b_{n} + \partial_a\partial_b\epsilon_n v^c_{n}\Big)
		= -\frac{1}{8}
        {\textstyle \sum_{n}}\partial_a(v^b_n v^c_n) f^{(1)}_n
	\end{aligned}
\end{equation}
Since the Drude contribution involves only intrinsic properties, this result constitutes a direct multiband extension of the two-band model.

\subsection{The \texorpdfstring{${\cal O}(\Gamma^{-1})$}{O(Gamma-1)} (Berry Curvature Dipole) Contribution}
The $\Gamma^{-1}$ contribution, $\sigma^{(-1)}_{abc}$, is derived as:
\begin{equation}
	\begin{aligned}
		\sigma^{(-1)}_{abc} &= -\frac{i}{4} 
        {\textstyle \sum_{nm}'}
        f^{(1)}_n \Big(v^c_{n}\frac{v^{a}_{mn}v^b_{nm} - v^{b}_{mn}v^a_{nm}}{\epsilon_{nm}^2} + v^b_{n}\frac{v^{a}_{mn}v^c_{nm} - v^{c}_{mn}v^a_{nm}}{\epsilon_{nm}^2} \Big) \\
	\end{aligned}
\end{equation}
By utilizing the multiband curvature-velocity relation, we can express this contribution in terms of the Berry curvature:
\begin{equation}
	\begin{aligned}
		\sigma^{(-1)}_{abc} 
		&= \frac{1}{4} {\textstyle \sum_{n}} f^{(1)}_n \Big(v^c_{n}\Omega^{ab}_n + v^b_{n}\Omega^{ac}_n \Big)
	\end{aligned}
\end{equation}
This result represents a natural generalization of the two-band case, achieved by replacing the two-band Berry curvature (defined via the anti-symmetric velocity product) with its multiband counterpart.

\subsection{The \texorpdfstring{${\cal O}(\Gamma^{0})$}{O(Gamma-0)} (Intrinsic) Contribution}
Following the calculation scheme in Supplementary Section \ref{Two-band Intrinsic}, we expand $\sigma^{(0)}_{abc}$ in powers of the inverse temperature $\beta$: 
\begin{equation}
\sigma^{(0)}_{abc} = T_0 + T_1 + T_2 + T_3 + T_4 ,
\end{equation}
which we investigate term by term below.

\subsubsection*{Term 4 (\texorpdfstring{$\sim \beta^4$}{~beta4}, \texorpdfstring{$f^{(4)}$}{f4})}
This contribution is proportional to the fourth derivative of Fermi-Dirac distribution:
\begin{equation}
	\begin{aligned}
		T_4 
		&= \frac{1}{4!} {\textstyle \sum_{n}} v^a_{n}v^{b}_{n}v^c_{n} f^{(4)}_n
        \label{s-rec-T4}
	\end{aligned}
\end{equation}

\subsubsection*{Term 3 (\texorpdfstring{$\sim \beta^3$}{beta3}, \texorpdfstring{$f^{(3)}$}{f3})}
This contribution is proportional to the third derivative of Fermi-Dirac distribution:
\begin{equation}
	\begin{aligned}
		T_3 
		&= {\textstyle \sum_{n}} \frac{f^{(3)}_n}{2\times 4!}  \Big[v^{ac}_{nn}v^b_{n} + v^{ab}_{nn}v^c_{n} + {\textstyle \sum_{m}'}\Big(v^b_{n}\frac{v^a_{nm}v^c_{mn}+v^c_{nm}v^a_{mn}}{\epsilon_{nm}} + v^c_{n}\frac{v^a_{nm}v^b_{mn}+v^b_{nm}v^a_{mn}}{\epsilon_{nm}}\Big) \Big] \\
	\end{aligned}
\end{equation}
Then, we utilize the property of Diagonal Second Derivative to simplify the expression:
\begin{equation}
	\begin{aligned}
		T_3 =  {\textstyle \sum_{n}} \frac{1}{2\times 4!}  \Big(\partial_a \partial_c \epsilon_n v^b_{n} + \partial_a \partial_b \epsilon_n v^c_{n}\Big)f^{(3)}_n
		= {\textstyle \sum_{n}}\frac{1}{2\times 4!}  \partial_a (v^c_{n} v^b_{n})f^{(3)}_n
	\end{aligned}
\end{equation}

\subsubsection*{Term 2 (\texorpdfstring{$\sim \beta^2$}{beta2}, \texorpdfstring{$f^{(2)}$}{f2})}
This contribution is proportional to the second derivative of Fermi-Dirac distribution:
\begin{equation}
	\begin{aligned}
		T_2 
		&= \frac{1}{4} {\textstyle \sum_{nm}'} f^{(2)}_n v^a_{n} \frac{v^b_{nm}v^c_{mn}+v^c_{nm}v^b_{mn}}{\epsilon^2_{nm}} \\
	\end{aligned}
\end{equation}
By using the Multiband Metric-Velocity Relation:
\begin{equation}
	T_2 = \frac{1}{2}{\textstyle \sum_{n}}f^{(2)}_n v^a_n g^{bc}_n
\end{equation}

\subsubsection*{Term 1 (\texorpdfstring{$\sim \beta^1$}{beta1}, \texorpdfstring{$f^{(1)}$}{f1})}
This contribution is proportional to the first derivative of Fermi-Dirac distribution:
\begin{equation}
	\begin{aligned}
		T_1
		&= {\textstyle \sum_{nm}'}\frac{1}{2} f^{(1)}_n \Big[v^b_{n} \frac{v^a_{nm}v^c_{mn} + v^c_{nm}v^a_{mn}}{\epsilon_{nm}^3}  + v^c_{n} \frac{v^a_{nm}v^b_{mn} + v^b_{nm}v^a_{mn}}{\epsilon_{nm}^3} - v^a_{n} \frac{v^b_{nm}v^c_{mn} + v^c_{nm}v^b_{mn}}{\epsilon_{nm}^3} \Big]
	\end{aligned}
\end{equation}
Applying the Multiband Band-renormalized Metric-Velocity Relation:
\begin{equation}
	T_1 
	= {\textstyle \sum_{n}} f^{(1)}_n \Big[v^b_{n} \mathcal{G}^{ac}_{n}  + v^c_{n} \mathcal{G}^{ab}_{n} - v^a_{n} \mathcal{G}^{bc}_{n} \Big]
    \label{multiband-T1}
\end{equation}

\subsubsection*{Term 0 (\texorpdfstring{$\sim \beta^0$}{beta0}, \texorpdfstring{$f^{(0)}$}{f0})}
Up to this point, our calculations for the multiband model show that terms $T_1$ through $T_4$ are consistent with the two-band model. While $T_0$ appears to introduce additional multiband contributions, we can demonstrate that these extra terms vanish. The explicit form of $T_0$ is:
\begin{equation}
	\begin{aligned}
		T_0
		&= {\textstyle \sum_{nm}'}\frac{1}{2}f_n\Big[3\Delta^a_{mn}\frac{v^{b}_{nm}v^{c}_{mn} + v^{c}_{nm}v^{b}_{mn}}{\epsilon_{nm}^4} +  \Delta^b_{mn}\frac{v^{a}_{nm}v^{c}_{mn} + v^{c}_{nm}v^{a}_{mn}}{\epsilon_{nm}^4}+ \Delta^c_{mn}\frac{v^{a}_{nm}v^{b}_{mn} + v^{b}_{nm}v^{a}_{mn}}{\epsilon_{nm}^4} \\
		&+ {\textstyle \sum_{l}'}\Big( \frac{v^a_{ln}v^b_{nm}v^c_{ml}}{\epsilon_{ln}\epsilon^3_{mn}} + \frac{v^a_{ln}v^c_{nm}v^b_{ml}}{\epsilon_{ln}\epsilon^3_{mn}} +
        \frac{v^a_{nl}v^c_{lm}v^b_{mn}}{\epsilon_{ln}\epsilon^3_{mn}} +
		\frac{v^a_{nl}v^b_{lm}v^c_{mn}}{\epsilon_{ln}\epsilon^3_{mn}}-\frac{v^a_{ml}v^b_{ln}v^c_{nm}}{\epsilon_{ml}\epsilon^3_{mn}} -\frac{v^a_{lm}v^b_{mn}v^c_{nl}}{\epsilon_{ml}\epsilon^3_{mn}}\\
		& -
		\frac{v^a_{ml}v^b_{nm}v^c_{ln}}{\epsilon_{ml}\epsilon^3_{mn}}  -
		\frac{v^a_{lm}v^b_{nl}v^c_{mn}}{\epsilon_{ml}\epsilon^3_{mn}}\Big)- \frac{v^{ac}_{nm}v^b_{mn}}{\epsilon^3_{mn}} -\frac{v^{ac}_{mn}v^b_{nm}}{\epsilon^3_{mn}} -\frac{v^{ab}_{nm}v^c_{mn}}{\epsilon^3_{mn}} -  \frac{v^{ab}_{mn}v^c_{nm}}{\epsilon^3_{mn}}\Big] \\ 
	\end{aligned}
\end{equation}
Applying the key identity Off-diagonal Second Derivative, we collect all multiband contributions (denoted as $\mathcal{M}$) arising from: (1) the explicit expression of $T_0$, and (2) the expansion of the off-diagonal second derivative.
\begin{equation}
	\begin{aligned}
		\mathcal{M} &= {\textstyle \sum_{nml}'} \frac{1}{2}f_n \Big[\Big( \frac{v^a_{ln}v^b_{nm}v^c_{ml}}{\epsilon_{ln}\epsilon^3_{mn}} + \frac{v^a_{ln}v^c_{nm}v^b_{ml}}{\epsilon_{ln}\epsilon^3_{mn}} +
		\frac{v^a_{nl}v^c_{lm}v^b_{mn}}{\epsilon_{ln}\epsilon^3_{mn}} +
		\frac{v^a_{nl}v^b_{lm}v^c_{mn}}{\epsilon_{ln}\epsilon^3_{mn}} \\
		& -\frac{v^a_{ml}v^b_{ln}v^c_{nm}}{\epsilon_{ml}\epsilon^3_{mn}} -\frac{v^a_{lm}v^b_{mn}v^c_{nl}}{\epsilon_{ml}\epsilon^3_{mn}}  -
		\frac{v^a_{ml}v^b_{nm}v^c_{ln}}{\epsilon_{ml}\epsilon^3_{mn}}  -
		\frac{v^a_{lm}v^b_{nl}v^c_{mn}}{\epsilon_{ml}\epsilon^3_{mn}}\Big) \\
		& + \Big(\frac{v^a_{nl}v^c_{lm}v^b_{mn}}{\epsilon_{nl}\epsilon_{mn}^3} -\frac{v^c_{nl}v^a_{lm}v^b_{mn}}{\epsilon_{lm}\epsilon_{mn}^3} 
		+\frac{v^a_{ml}v^c_{ln}v^b_{nm}}{\epsilon_{ml}\epsilon_{mn}^3} -\frac{v^c_{ml}v^a_{ln}v^b_{nm}}{\epsilon_{ln}\epsilon_{mn}^3} \\
		& + \frac{v^a_{nl}v^b_{lm}v^c_{mn}}{\epsilon_{nl}\epsilon_{mn}^3} -\frac{v^b_{nl}v^a_{lm}v^c_{mn}}{\epsilon_{lm}\epsilon_{mn}^3} 
		+\frac{v^a_{ml}v^b_{ln}v^c_{nm}}{\epsilon_{ml}\epsilon_{mn}^3} -\frac{v^b_{ml}v^a_{ln}v^c_{nm}}{\epsilon_{ln}\epsilon_{mn}^3} \Big)\Big] = 0
	\end{aligned}
\end{equation}

Consequently, only terms involving two-band indices contribute:
\begin{equation}
	\begin{aligned}
		T_0
		&= {\textstyle \sum_{nm}'} \frac{1}{2}f_n\Big[3\Delta^a_{mn}\frac{v^{b}_{nm}v^{c}_{mn} + v^{c}_{nm}v^{b}_{mn}}{\epsilon_{nm}^4} +  \Delta^b_{mn}\frac{v^{a}_{nm}v^{c}_{mn} + v^{c}_{nm}v^{a}_{mn}}{\epsilon_{nm}^4}+ \Delta^c_{mn}\frac{v^{a}_{nm}v^{b}_{mn} + v^{b}_{nm}v^{a}_{mn}}{\epsilon_{nm}^4} \\
		& - \frac{1}{\epsilon^3_{mn}} \Big(-i R^{ac}_{nm}v^{c}_{nm} v^{b}_{mn} + \frac{v^c_{nm}v^b_{mn} \Delta_{nm}^a}{\epsilon_{nm}} + \frac{v^a_{nm}v^b_{mn} \Delta_{nm}^c}{\epsilon_{nm}} -i R^{ac}_{mn}v^{c}_{mn} v^{b}_{nm} + \frac{v^c_{mn}v^b_{nm} \Delta_{mn}^a}{\epsilon_{mn}} + \frac{v^a_{mn}v^b_{nm} \Delta_{mn}^c}{\epsilon_{mn}}\\
		& -i R^{ab}_{nm}v^{b}_{nm} v^{c}_{mn} + \frac{v^b_{nm}v^c_{mn} \Delta_{nm}^a}{\epsilon_{nm}} + \frac{v^a_{nm}v^c_{mn} \Delta_{nm}^b}{\epsilon_{nm}} -i R^{ab}_{mn}v^{b}_{mn} v^{c}_{nm}+ \frac{v^b_{mn}v^c_{nm} \Delta_{mn}^a}{\epsilon_{mn}} + \frac{v^a_{mn}v^c_{nm} \Delta_{mn}^b}{\epsilon_{mn}}\Big)\Big]
	\end{aligned}
\end{equation}
We utilize the properties of shift vector $-i(R^{ac}_{nm} + R^{ab}_{mn})v^c_{nm}v^{b}_{mn} = \epsilon^2_{nm} \partial_a (A^c_{nm}A^b_{mn})$:
\begin{equation}
	\begin{aligned}
		T_0
		&= {\textstyle \sum_{nm}'}\frac{1}{2}f_n\Big[ \frac{\partial_a (A^{c}_{nm}A^{b}_{mn} + A^{b}_{nm}A^{c}_{mn})}{\epsilon_{nm}} +\Delta^a_{mn}\frac{v^{b}_{nm}v^{c}_{mn} + v^{c}_{nm}v^{b}_{mn}}{\epsilon_{nm}^4}\Big] \\
	\end{aligned}
\end{equation}

By further applying the Feynman-Hellmann identity and the definition of the band-renormalized quantum metric, the final expression is shown to be proportional to the derivative of the band-renormalized quantum metric:
\begin{equation}
	\begin{aligned}
		T_0
		&= {\textstyle \sum_{nm}'} \frac{1}{2}f_n\Big[ \frac{\partial_a (A^{c}_{nm}A^{b}_{mn} + A^{b}_{nm}A^{c}_{mn})}{\epsilon_{nm}} + (A^{b}_{nm}A^{c}_{mn} + A^{c}_{nm}A^{b}_{mn})\partial_a(\frac{1}{\epsilon_{nm}})\Big] \\
		&= {\textstyle \sum_{nm}'} \frac{1}{2}f_n \partial_a \Big(\frac{A^{c}_{nm}A^{b}_{mn} + A^{b}_{nm}A^{c}_{mn}}{\epsilon_{nm}}\Big) = {\textstyle \sum_{n}}f_n \partial_a \mathcal{G}^{bc}_{n} \\
	\end{aligned}
\end{equation}

\subsection{Transformation and Final Expressions}
We employ IBP in $k$-space to simplify the expressions and facilitate a comparison with the two-band model results presented in Supplementary Section \ref{Two-band Final Expression}. Our derivation reveals that the final expressions of $\sigma^{(0)}_{abc}$ represent a natural generalization of the two-band model; the analytical form remains identical, requiring (1) the extension of the band summation index from $n \in \{1, 2\}$ to $n \in \{1, \dots, N\}$ (2) replacing the two-band (Band-renormalized) Quantum Metric with their multiband counterpart. The detailed results of this derivation are summarized below:

\subsubsection*{A. Kinetic Contribution}
The total intrinsic kinetic contribution is $\sigma^{\text{kin}}_{abc} = T_4 + T_3$.  By performing IBP on $T_3$ with respect to $k_a$, we obtain:
\begin{equation}
	T_3 \xrightarrow{\text{IBP}} -{\textstyle \sum_{n}} \frac{1}{2\times 4!}  (v^a_{n} v^c_{n} v^b_{n})f^{(4)}_n = -\frac{1}{2}T_4
\end{equation}

Summing the resulting terms yields:
\begin{equation}
	\sigma^{\text{kin}}_{abc} = \frac{1}{2}T_4 = {\textstyle \sum_{n}}\frac{1}{48} v^{a}_{n}v^{b}_{n}v^c_{n}f^{(4)}_{n},
    \quad
    f_n^{(4)} = 2 f_{0,n}^{(4)}
\end{equation}

\subsubsection*{B. Geometric Contribution}
The rest of intrinsic contribution is geometric $\sigma^{\text{geo}}_{abc} = T_2 + T_1 + T_0$. First, we apply IBP to $T_0$:
\begin{equation}
	T_0 \xrightarrow{\text{IBP}} - {\textstyle \sum_{n}} f^{(1)}_n v_n^a \mathcal{G}^{bc}_{n} 
\end{equation}

Next, performing IBP on $T_2$ leads to:
\begin{equation}
	T_2 \xrightarrow{\text{IBP}}-\frac{1}{2} {\textstyle \sum_{n}}f^{(1)}_n \partial_{a}g^{bc}_n
\end{equation}

Combining the above contributions with $T_1$ [Eq.~\eqref{multiband-T1}], we arrive at the final expression:
\begin{equation}
	\begin{aligned}
		\sigma^{\text{geo}}_{abc} &= -\frac{1}{2} {\textstyle \sum_{n}}f^{(1)}_n \partial_{a}g^{bc}_n  + {\textstyle \sum_{n}} f^{(1)}_n \Big[v^b_{n} \mathcal{G}^{ac}_{n}  + v^c_{n} \mathcal{G}^{ab}_{n} - v^a_{n} \mathcal{G}^{bc}_{n} \Big] -{\textstyle \sum_{n}} f^{(1)}_{n} v^a_{n} \mathcal{G}^{bc}_n \\
		&= {\textstyle \sum_{n}} f^{(1)}_n \Big[v^b_{n} \mathcal{G}^{ac}_{n}  + v^c_{n} \mathcal{G}^{ab}_{n} - 2v^a_{n} \mathcal{G}^{bc}_{n}  - \frac{\partial_a g^{bc}_n}{2}\Big],
        \quad
        f_n^{(1)} = 2 f_{0,n}^{(1)}
	\end{aligned}
\end{equation}
This expression closely mirrors its two-band counterpart (Eq. \ref{Two-band Geometric}), yet entails a multiband generalization: the term $g_{ab}/\epsilon_{n\bar{n}}$ is now replaced by (or generalized to) the multiband band-renormalized quantum metric $\mathcal{G}^{ab}_n$.

\section{DC Conductivity from the Low-Frequency Limit of Second-Harmonic Generation (SHG)}
In experimental practice, the low-frequency SHG conductivity, $\sigma_{abc}(-2\omega; \omega, \omega)$, is often preferred over the rectified conductivity, $\sigma_{abc}(\omega, -\omega)$. Typical experimental frequencies range from $10^1$ to $10^3$ Hz, which are significantly lower than the scattering rates in most materials. Consequently, these measurements can be effectively treated as the low-frequency limit ($\omega \rightarrow 0$) of the SHG response. To better align our theoretical framework with experimental observations, this section elucidates the SHG response in this limit.

We adopt a procedure analogous to the calculation of the rectified conductivity illustrated in Supplementary Section \ref{S-AA}, \ref{S-B} and \ref{S-C}: specifically, we evaluate the low-frequency limit of the response function to extract the frequency-independent nonlinear transport coefficients, and subsequently perform a relaxation-time expansion around $\Gamma = 0$:
\begin{equation}
	\sigma_{abc}^{\text{SHG}} = \frac{1}{\Gamma^2}\sigma_{abc}^{\text{SHG}(-2)} + \frac{1}{\Gamma}\sigma_{abc}^{\text{SHG}(-1)} + \sigma_{abc}^{\text{SHG}(0)} + \mathcal{O}(\Gamma)
\end{equation}

\subsection{The \texorpdfstring{${\cal O}(\Gamma^{-2})$}{O(Gamma-2)} (Nonlinear Drude) Contribution}
The $\Gamma^{-2}$ contribution is given by:
\begin{equation}
	\begin{aligned}
		\sigma^{\text{SHG}(-2)}_{abc} &=  {\textstyle \sum_{n}}\frac{1}{2}f^{(2)}_n v^a_n v^b_n v^c_n + \frac{1}{8} {\textstyle \sum_{n}} f^{(1)}_{n} \Big[ v^{ac}_{nn}v^b_{n} + v^{ab}_{nn}v^c_{n} + 4v^{bc}_{nn}v^a_{n}\\
		& + {\textstyle \sum_{m}'} \Big( v^c_{n}\frac{v^{a}_{mn}v^b_{nm} + v^{b}_{mn}v^a_{nm}}{\epsilon_{nm}} + v^b_{n}\frac{v^{a}_{mn}v^c_{nm} + v^{c}_{mn}v^a_{nm}}{\epsilon_{nm}} + 4v^a_{n}\frac{v^{b}_{mn}v^c_{nm} + v^{c}_{mn}v^b_{nm}}{\epsilon_{nm}}\Big) \Big] \\
	\end{aligned}
\end{equation}

By applying the properties of the diagonal second derivative (as detailed in Supplementary Section \ref{S-E}) and performing integration by parts (IBP), we obtain:
\begin{equation}
	\begin{aligned}
		\sigma^{\text{SHG}(-2)}_{abc} 
		&= {\textstyle \sum_{n}}\Big[\frac{1}{2}f^{(2)}_n v^a_n v^b_n v^c_n + \frac{1}{8} f^{(1)}_{n} \Big(\partial_a\partial_c \epsilon_n v^b_n + \partial_a\partial_b \epsilon_n v^c_n + 4\partial_b\partial_c \epsilon_n v^a_n\Big)\Big] \\
		&\xrightarrow{\text{IBP}}  -\frac{1}{8}{\textstyle \sum_{n}} f^{(1)}_n \partial_a(v^b_n v^c_n)
	\end{aligned}
\end{equation}
Consequently, we find that in the low-frequency limit, the nonlinear Drude contribution to the SHG is identical to the corresponding expression for the rectified conductivity.

\subsection{The \texorpdfstring{${\cal O}(\Gamma^{-1})$}{O(Gamma-1)} (Berry Curvature Dipole) Contribution}
The $\Gamma^{-1}$ contribution is expressed as:
\begin{equation}
	\begin{aligned}
		\sigma^{\text{SHG}(-1)}_{abc} &= -\frac{i}{4} {\textstyle \sum_{nm}'} f^{(1)}_n \Big(v^c_{n}\frac{v^{a}_{mn}v^b_{nm} - v^{b}_{mn}v^a_{nm}}{\epsilon_{nm}^2} + v^b_{n}\frac{v^{a}_{mn}v^c_{nm} - v^{c}_{mn}v^a_{nm}}{\epsilon_{nm}^2} \Big) \\
	\end{aligned}
\end{equation}
Utilizing the multiband curvature-velocity relation:
\begin{equation}
	\begin{aligned}
		\sigma^{\text{SHG}(-1)}_{abc}
		&= \frac{1}{4} {\textstyle \sum_{n}} f^{(1)}_n \Big(v^c_{n}\Omega^{ab}_n + v^b_{n}\Omega^{ac}_n \Big)
	\end{aligned}
\end{equation}
From this result, it is evident that the $\Gamma^{-1}$ expansion of the SHG is directly related to the Berry curvature, and its expression is identical to that of the rectified conductivity.

\subsection{The \texorpdfstring{${\cal O}(\Gamma^{0})$}{O(Gamma-0)} (Intrinsic) Contribution}
Analogous to the rectified conductivity, the SHG $\Gamma^0$ term $\sigma_{abc}^{\text{SHG}(0)}$ is expanded in powers of the inverse temperature $\beta$ as follows:
\begin{equation}
	\sigma_{abc}^{\text{SHG}(0)} = T^{\text{SHG}}_0 + T^{\text{SHG}}_1 + T^{\text{SHG}}_2 + T^{\text{SHG}}_3 + T^{\text{SHG}}_4.
\end{equation}

\subsubsection*{Term 4 (\texorpdfstring{$\sim \beta^4$}{~beta4}, \texorpdfstring{$f^{(4)}$}{f4})}
The fourth-order SHG contribution ($T^{\text{SHG}}_4$) is the negative of the one of the rectified current [$T_4$, Eq.~\eqref{s-rec-T4}]:
\begin{equation}
	T^{\text{SHG}}_4 = -T_4 = -\frac{1}{4!} {\textstyle \sum_{n}} v^a_{n} v^{b}_{n}v^c_{n} f^{(4)}_n
\end{equation}

\subsubsection*{Term 3 (\texorpdfstring{$\sim \beta^3$}{beta3}, \texorpdfstring{$f^{(3)}$}{f3})}
The third-order contribution to the SHG response, $T^{\text{SHG}}_3$, is expressed as:
\begin{equation}
	\begin{aligned}
		T^{\text{SHG}}_3 &= -\frac{1}{2 \times 4!} {\textstyle \sum_{n}} f^{(3)}_n \Big[v^{ac}_{nn} v^b_{n} + v^{ab}_{nn}v^c_{n} + 4v^{bc}_{nn}v^a_{n} \\
		&  + {\textstyle \sum_{m}'}\Big(v^c_{n}\frac{v^a_{nm}v^b_{mn} + v^b_{nm}v^a_{mn}}{\epsilon_{nm}} + v^b_{n}\frac{v^a_{nm}v^c_{mn} + v^c_{nm}v^a_{mn}}{\epsilon_{nm}} + 4v^a_{n}\frac{v^b_{nm}v^c_{mn} + v^c_{nm}v^b_{mn}}{\epsilon_{nm}}\Big)\Big] \\
	\end{aligned}
\end{equation}
By utilizing the properties of the Diagonal Second Derivative, we can simplify the expression to:
\begin{equation}
	\begin{aligned}
		T^{\text{SHG}}_3 
		&= -\frac{1}{2 \times 4!} {\textstyle \sum_{n}} f^{(3)}_n \Big[\partial_a\partial_c \epsilon_n \partial_b \epsilon_n + \partial_a\partial_b \epsilon_n \partial_c \epsilon_n + 4\partial_b\partial_c \epsilon_n \partial_a \epsilon_n\Big] \\
	\end{aligned}
\end{equation}

\subsubsection*{Term 2 (\texorpdfstring{$\sim \beta^2$}{beta2}, \texorpdfstring{$f^{(2)}$}{f2})}
The second-order SHG term $T^{\text{SHG}}_2$ is found to be identical to its counterpart in the rectified current:
\begin{equation}
	\begin{aligned}
		T^{\text{SHG}}_2 = \frac{1}{2}{\textstyle \sum_{n}} f^{(2)}_n v^{a}_ng^{bc}_n
	\end{aligned}
\end{equation}

\subsubsection*{Term 1 (\texorpdfstring{$\sim \beta^1$}{beta1}, \texorpdfstring{$f^{(1)}$}{f1})}
The first-order SHG term $T^{\text{SHG}}_1$ is given by:
\begin{equation}
	\begin{aligned}
		T^{\text{SHG}}_1 &= - {\textstyle \sum_{nm}'}\frac{1}{2} f^{(1)}_n \Big[v^a_{n} \frac{v^b_{nm}v^c_{mn} + v^c_{nm}v^b_{mn}}{\epsilon_{nm}^3} + v^b_{n} \frac{v^a_{nm}v^c_{mn} + v^c_{nm}v^a_{mn}}{\epsilon_{nm}^3}  + v^c_{n} \frac{v^a_{nm}v^b_{mn} + v^b_{nm}v^a_{mn}}{\epsilon_{nm}^3} \Big] 
	\end{aligned}
\end{equation}
By invoking the multiband band-renormalized metric-velocity relation, this expression can be simplified as follows:
\begin{equation}
	\begin{aligned}
		T^{\text{SHG}}_1
		&= - {\textstyle \sum_{n}} f^{(1)}_n \Big[v^b_{n} \mathcal{G}^{ac}_{n}  + v^c_{n} \mathcal{G}^{ab}_{n} + v^a_{n} \mathcal{G}^{bc}_{n} \Big]
	\end{aligned}
\end{equation}

\subsubsection*{Term 0 (\texorpdfstring{$\sim \beta^0$}{beta0}, \texorpdfstring{$f^{(0)}$}{f0})}
	The simplification of the SHG term $T^{\text{SHG}}_0$ is the most analytically involved step. For clarity, we decompose the initial expression into five distinct components:
\begin{equation}
	T^{\text{SHG}}_{0, 1} = \frac{1}{2} {\textstyle \sum_{nm}'} f_n  \Big[\frac{\Delta_{nm}^a (v^b_{mn}v^c_{nm} + v^c_{mn}v^b_{nm})}{\epsilon^4_{nm}} + 7\frac{\Delta_{nm}^c (v^a_{mn}v^b_{nm} + v^b_{mn}v^a_{nm})}{\epsilon^4_{nm}} + 7\frac{\Delta_{nm}^b (v^a_{mn}v^c_{nm} + v^c_{mn}v^a_{nm})}{\epsilon^4_{nm}}\Big]
\end{equation}
\begin{equation}
	T^{\text{SHG}}_{0, 2} = -\frac{1}{2} {\textstyle \sum_{nm}'} f_n  \Big(\frac{v^{ac}_{nm}v^b_{mn}}{\epsilon_{nm}^3} + \frac{v^{ab}_{nm}v^c_{mn}}{\epsilon_{nm}^3} + \frac{v^{ac}_{mn}v^b_{nm}}{\epsilon_{nm}^3} + 
	\frac{v^{ab}_{mn}v^c_{nm}}{\epsilon_{nm}^3} + 
	\frac{4v^{bc}_{nm}v^a_{mn}}{\epsilon_{nm}^3} + 
	\frac{4v^{bc}_{mn}v^a_{nm}}{\epsilon_{nm}^3}\Big) 
\end{equation}
\begin{equation}
	\begin{aligned}
		T^{\text{SHG}}_{0, 3} &= -\frac{1}{2} {\textstyle \sum_{nml}'} f_n \Big( \frac{v^{a}_{nl}v^b_{mn}v^c_{lm}}{\epsilon_{nl}\epsilon^3_{nm}} + \frac{v^{a}_{nl}v^b_{lm}v^c_{mn}}{\epsilon_{nl}\epsilon^3_{nm}} + 
		\frac{v^{a}_{ln}v^b_{ml}v^c_{nm}}{\epsilon_{nl}\epsilon^3_{nm}} + 
		\frac{v^{a}_{ln}v^b_{nm}v^c_{ml}}{\epsilon_{nl}\epsilon^3_{nm}}  \\
		&- \frac{v^{a}_{lm}v^b_{mn}v^c_{nl}}{\epsilon_{lm}\epsilon^3_{nm}} - \frac{v^{a}_{lm}v^b_{nl}v^c_{mn}}{\epsilon_{lm}\epsilon^3_{nm}} - 
		\frac{v^{a}_{ml}v^b_{ln}v^c_{nm}}{\epsilon_{lm}\epsilon^3_{nm}} - 
		\frac{v^{a}_{ml}v^b_{nm}v^c_{ln}}{\epsilon_{lm}\epsilon^3_{nm}}  \Big)
	\end{aligned}
\end{equation}
\begin{equation}
	\begin{aligned}
		T^{\text{SHG}}_{0, 4} &=   - {\textstyle \sum_{nml}'} f_n \Big( \frac{v^{a}_{nl}v^b_{mn}v^c_{lm}}{\epsilon^2_{nl}\epsilon^2_{nm}} + \frac{v^{a}_{nl}v^b_{lm}v^c_{mn}}{\epsilon^2_{nl}\epsilon^2_{nm}} + 
		\frac{v^{a}_{ln}v^b_{ml}v^c_{nm}}{\epsilon^2_{nl}\epsilon^2_{nm}} + 
		\frac{v^{a}_{ln}v^b_{nm}v^c_{ml}}{\epsilon^2_{nl}\epsilon^2_{nm}}  \\
		&- \frac{v^{a}_{lm}v^b_{mn}v^c_{nl}}{\epsilon^2_{lm}\epsilon^2_{nm}} - \frac{v^{a}_{lm}v^b_{nl}v^c_{mn}}{\epsilon^2_{lm}\epsilon^2_{nm}} - 
		\frac{v^{a}_{ml}v^b_{ln}v^c_{nm}}{\epsilon^2_{lm}\epsilon^2_{nm}} - 
		\frac{v^{a}_{ml}v^b_{nm}v^c_{ln}}{\epsilon^2_{lm}\epsilon^2_{nm}}  \Big)
	\end{aligned}
\end{equation}
\begin{equation}
	\begin{aligned}
		T^{\text{SHG}}_{0, 5} &= -2 {\textstyle \sum_{nml}'} f_n \Big( \frac{v^{a}_{nl}v^b_{mn}v^c_{lm}}{\epsilon^3_{nl}\epsilon_{nm}} + \frac{v^{a}_{nl}v^b_{lm}v^c_{mn}}{\epsilon^3_{nl}\epsilon_{nm}} + 
		\frac{v^{a}_{ln}v^b_{ml}v^c_{nm}}{\epsilon^3_{nl}\epsilon_{nm}} + 
		\frac{v^{a}_{ln}v^b_{nm}v^c_{ml}}{\epsilon^3_{nl}\epsilon_{nm}}  \\
		&- \frac{v^{a}_{lm}v^b_{mn}v^c_{nl}}{\epsilon^3_{lm}\epsilon_{ln}} - \frac{v^{a}_{lm}v^b_{nl}v^c_{mn}}{\epsilon^3_{lm}\epsilon_{ln}} - 
		\frac{v^{a}_{ml}v^b_{ln}v^c_{nm}}{\epsilon^3_{lm}\epsilon_{ln}} - 
		\frac{v^{a}_{ml}v^b_{nm}v^c_{ln}}{\epsilon^3_{lm}\epsilon_{ln}}  \Big)
	\end{aligned}
\end{equation}
The terms $T^{\text{SHG}}_{0, 3}$, $T^{\text{SHG}}_{0, 4}$, and $T^{\text{SHG}}_{0, 5}$ explicitly contain multiband contributions. Meanwhile, $T^{\text{SHG}}_{0, 2}$ involves the Off-diagonal Second Derivative; based on our analysis in Section \ref{S-E}, expanding this term using the properties of Off-diagonal Second Derivatives will yield a series of additional multiband contributions.

Starting from $T^{\text{SHG}}_{0, 2}$, we expand the Off-diagonal Second Derivatives first and decompose the shift vector into two distinct components: the logarithmic derivative $i\partial_{a} (\ln A^{b}_{nm})$ and the difference in intra-band Berry connections $(A^{a}_{nn} - A^{a}_{mm})$.
\begin{equation}
	\begin{aligned}
		&T^{\text{SHG}}_{0, 2} =- \frac{1}{2} {\textstyle \sum_{nm}'} \frac{f_n}{\epsilon^3_{nm}}\Big[\partial_a(\ln A^{c}_{nm}) v^c_{nm} v^{b}_{mn} + \partial_a(\ln A^{b}_{nm}) v^b_{nm} v^{c}_{mn} + \partial_a(\ln A^{c}_{mn}) v^c_{mn} v^{b}_{nm} + \partial_a(\ln A^{b}_{mn}) v^b_{mn} v^{c}_{nm} \\
		&+ 2\partial_c(\ln A^{b}_{nm}) v^b_{nm} v^{a}_{mn} + 2\partial_c(\ln A^{b}_{mn}) v^b_{mn} v^{a}_{nm} + 2\partial_b(\ln A^{c}_{nm}) v^c_{nm} v^{a}_{mn} + 2\partial_b(\ln A^{c}_{mn}) v^c_{mn} v^{a}_{nm}\\
		& -2i (A^c_{nn} - A^c_{mm})v^b_{nm} v^{a}_{mn}-2i (A^c_{mm} - A^c_{nn})v^b_{mn} v^{a}_{nm} -2i (A^b_{nn} - A^b_{mm})v^c_{nm} v^{a}_{mn}-2i (A^b_{mm} - A^b_{nn})v^c_{mn} v^{a}_{nm}\\
		& + 2\frac{\Delta_{nm}^a (v^b_{mn}v^c_{nm} + v^c_{mn}v^b_{nm})}{\epsilon_{nm}} + 5\frac{\Delta_{nm}^c (v^a_{mn}v^b_{nm} + v^b_{mn}v^a_{nm})}{\epsilon_{nm}} + 5\frac{\Delta_{nm}^b (v^a_{mn}v^c_{nm} + v^c_{mn}v^a_{nm})}{\epsilon_{nm}} \\
		&+ {\textstyle \sum_{l}'} \Big(
		-2\frac{v^b_{nl}v^c_{lm}v^a_{mn}}{\epsilon_{nl}} - 2\frac{v^c_{nl}v^b_{lm}v^a_{mn}}{\epsilon_{nl}} -
		2\frac{v^c_{ml}v^b_{ln}v^a_{nm}}{\epsilon_{nl}} - 
		2\frac{v^b_{ml}v^c_{ln}v^a_{nm}}{\epsilon_{nl}} 
        + 2\frac{v^c_{nl}v^b_{lm}v^a_{mn}}{\epsilon_{lm}} + 2\frac{v^b_{nl}v^c_{lm}v^a_{mn}}{\epsilon_{lm}} +
		2\frac{v^b_{ml}v^c_{ln}v^a_{nm}}{\epsilon_{lm}} + 
		2\frac{v^c_{ml}v^b_{ln}v^a_{nm}}{\epsilon_{lm}} \\
		&-\frac{v^{a}_{nl}v^b_{mn}v^c_{lm}}{\epsilon_{nl}} - \frac{v^{a}_{nl}v^b_{lm}v^c_{mn}}{\epsilon_{nl}} -
		\frac{v^{a}_{ln}v^b_{ml}v^c_{nm}}{\epsilon_{nl}} - 
		\frac{v^{a}_{ln}v^b_{nm}v^c_{ml}}{\epsilon_{nl}}
        + \frac{v^{a}_{lm}v^b_{mn}v^c_{nl}}{\epsilon_{lm}} + \frac{v^{a}_{lm}v^b_{nl}v^c_{mn}}{\epsilon_{lm}} +
		\frac{v^{a}_{ml}v^b_{ln}v^c_{nm}}{\epsilon_{lm}} + 
		\frac{v^{a}_{ml}v^b_{nm}v^c_{ln}}{\epsilon_{lm}}\Big)\Big]\\
	\end{aligned}
\end{equation}
We first notice the last two lines cancel $T^{\text{SHG}}_{0, 3}$. Then we try to rewrite Berry-connection-difference part like $(A^c_{nn} - A^c_{mm})v^b_{nm} v^{a}_{mn}$ by using properties of off-diagonal second derivative again:
\begin{equation}
	\begin{aligned}
		&-i(A^c_{mm} - A^c_{nn})v^b_{mn} v^{a}_{nm} -i (A^b_{nn} - A^b_{mm})v^c_{nm} v^{a}_{mn} \\
		&= -\partial_a(\ln A^{c}_{nm}) v^c_{nm} v^{b}_{mn} - \partial_a(\ln A^{b}_{mn}) v^b_{mn} v^{c}_{nm} + \partial_c(\ln A^{a}_{nm}) v^a_{nm} v^{b}_{mn} + \partial_b(\ln A^{a}_{mn}) v^a_{mn} v^{c}_{nm} \\
		& + {\textstyle \sum_{l}'} \Big( -\frac{v^c_{nl}v^a_{lm}v^b_{mn}}{\epsilon_{nl}} + \frac{v^a_{nl}v^c_{lm}v^b_{mn}}{\epsilon_{lm}} - \frac{v^b_{ml}v^a_{ln}v^c_{nm}}{\epsilon_{ml}}+ \frac{v^a_{ml}v^b_{ln}v^c_{nm}}{\epsilon_{ln}} 
        + \frac{v^a_{nl}v^c_{lm}v^b_{mn}}{\epsilon_{nl}} - \frac{v^c_{nl}v^a_{lm}v^b_{mn}}{\epsilon_{lm}} + \frac{v^a_{ml}v^b_{ln}v^c_{nm}}{\epsilon_{ml}}- \frac{v^b_{ml}v^a_{ln}v^c_{nm}}{\epsilon_{ln}}\Big)
	\end{aligned}
\end{equation}

By combining $T^{\text{SHG}}_{0, 2}$ and $T^{\text{SHG}}_{0, 3}$, we can decompose the resulting expression into a two-band term, $T^{\text{SHG; two-band}}_{0, 2+3}$, and a multiband contribution, $T^{\text{SHG; multiband}}_{0, 2+3}$:
\begin{equation}
	\begin{aligned}
		T^{\text{SHG}}_{0, 2+3} &= T^{\text{SHG}}_{0, 2} + T^{\text{SHG}}_{0, 3} \\
		&=- \frac{1}{2} {\textstyle \sum_{nm}'} \frac{f_n}{\epsilon^3_{nm}}\Big[\partial_a(\ln A^{c}_{nm}) v^c_{nm} v^{b}_{mn} + \partial_a(\ln A^{b}_{nm}) v^b_{nm} v^{c}_{mn} + \partial_a(\ln A^{c}_{mn}) v^c_{mn} v^{b}_{nm} + \partial_a(\ln A^{b}_{mn}) v^b_{mn} v^{c}_{nm} \\
		&+ 2\partial_c(\ln A^{b}_{nm}) v^b_{nm} v^{a}_{mn} + 2\partial_c(\ln A^{b}_{mn}) v^b_{mn} v^{a}_{nm} + 2\partial_b(\ln A^{c}_{nm}) v^c_{nm} v^{a}_{mn} + 2\partial_b(\ln A^{c}_{mn}) v^c_{mn} v^{a}_{nm}\\
		&+2\partial_c(\ln A^{a}_{mn}) v^a_{mn} v^{b}_{nm} + 2\partial_c(\ln A^{a}_{nm}) v^a_{nm} v^{b}_{mn} + 2\partial_b(\ln A^{a}_{mn}) v^a_{mn} v^{c}_{nm} + 2\partial_b(\ln A^{a}_{nm}) v^a_{nm} v^{c}_{mn} \\
		&-2\partial_a(\ln A^{c}_{nm}) v^c_{nm} v^{b}_{mn} - 2\partial_a(\ln A^{b}_{nm}) v^b_{nm} v^{c}_{mn} - 2\partial_a(\ln A^{c}_{mn}) v^c_{mn} v^{b}_{nm} - 2\partial_a(\ln A^{b}_{mn}) v^b_{mn} v^{c}_{nm} \\
		& + 2\frac{\Delta_{nm}^a (v^b_{mn}v^c_{nm} + v^c_{mn}v^b_{nm})}{\epsilon_{nm}} + 5\frac{\Delta_{nm}^c (v^a_{mn}v^b_{nm} + v^b_{mn}v^a_{nm})}{\epsilon_{nm}} + 5\frac{\Delta_{nm}^b (v^a_{mn}v^c_{nm} + v^c_{mn}v^a_{nm})}{\epsilon_{nm}} \\
		&+ {\textstyle \sum_{l}'} \Big(-2\frac{v^b_{nl}v^c_{lm}v^a_{mn}}{\epsilon_{nl}} - 2\frac{v^c_{nl}v^b_{lm}v^a_{mn}}{\epsilon_{nl}} -
		2\frac{v^c_{ml}v^b_{ln}v^a_{nm}}{\epsilon_{nl}} - 
		2\frac{v^b_{ml}v^c_{ln}v^a_{nm}}{\epsilon_{nl}}  \\
		&+ 2\frac{v^c_{nl}v^b_{lm}v^a_{mn}}{\epsilon_{lm}} + 2\frac{v^b_{nl}v^c_{lm}v^a_{mn}}{\epsilon_{lm}} +
		2\frac{v^b_{ml}v^c_{ln}v^a_{nm}}{\epsilon_{lm}} + 
		2\frac{v^c_{ml}v^b_{ln}v^a_{nm}}{\epsilon_{lm}} \\
		&+ 2\frac{v^a_{ml}v^c_{ln}v^b_{nm}}{\epsilon_{ml}} + 2\frac{v^a_{ml}v^b_{ln}v^c_{nm}}{\epsilon_{ml}} + 2\frac{v^a_{nl}v^c_{lm}v^b_{mn}}{\epsilon_{nl}} +
		2\frac{v^a_{nl}v^b_{lm}v^c_{mn}}{\epsilon_{nl}} \\
		&- 2\frac{v^c_{ml}v^a_{ln}v^b_{nm}}{\epsilon_{ln}} - 2\frac{v^b_{ml}v^a_{ln}v^c_{nm}}{\epsilon_{ln}} - 2\frac{v^c_{nl}v^a_{lm}v^b_{mn}}{\epsilon_{lm}} -
		2\frac{v^b_{nl}v^a_{lm}v^c_{mn}}{\epsilon_{lm}} \\
		&- 2\frac{v^c_{ml}v^a_{ln}v^b_{nm}}{\epsilon_{ml}} - 2\frac{v^b_{ml}v^a_{ln}v^c_{nm}}{\epsilon_{ml}} - 2\frac{v^c_{nl}v^a_{lm}v^b_{mn}}{\epsilon_{nl}} -
		2\frac{v^b_{nl}v^a_{lm}v^c_{mn}}{\epsilon_{nl}} \\
		&+ 2\frac{v^a_{ml}v^c_{ln}v^b_{nm}}{\epsilon_{ln}} + 2\frac{v^a_{ml}v^b_{ln}v^c_{nm}}{\epsilon_{ln}} + 2\frac{v^a_{nl}v^c_{lm}v^b_{mn}}{\epsilon_{lm}} +
		2\frac{v^a_{nl}v^b_{lm}v^c_{mn}}{\epsilon_{lm}} \Big)\Big]
        = T^{\text{SHG; two-band}}_{0, 2+3} + T^{\text{SHG; multiband} }_{0, 2+3}
	\end{aligned}
\end{equation}

At this stage, we shall now group the multiband contributions:
\begin{equation}
	\begin{aligned}
		&\mathcal{M} = T^{\text{SHG; multiband} }_{0, 2+3} + T^{\text{SHG}}_{0, 4} + T^{\text{SHG}}_{0, 5} \\
		&= - \frac{1}{2} {\textstyle \sum_{nml}'} f_n\Big[\frac{2v^{a}_{nl}v^b_{mn}v^c_{lm}}{\epsilon^2_{nl}\epsilon^2_{nm}} + \frac{2v^{a}_{nl}v^b_{lm}v^c_{mn}}{\epsilon^2_{nl}\epsilon^2_{nm}} + 
		\frac{2v^{a}_{ln}v^b_{ml}v^c_{nm}}{\epsilon^2_{nl}\epsilon^2_{nm}} + 
		\frac{2v^{a}_{ln}v^b_{nm}v^c_{ml}}{\epsilon^2_{nl}\epsilon^2_{nm}}  \\
		&- \frac{2v^{a}_{lm}v^b_{mn}v^c_{nl}}{\epsilon^2_{lm}\epsilon^2_{nm}} - \frac{2v^{a}_{lm}v^b_{nl}v^c_{mn}}{\epsilon^2_{lm}\epsilon^2_{nm}} - 
		\frac{2v^{a}_{ml}v^b_{ln}v^c_{nm}}{\epsilon^2_{lm}\epsilon^2_{nm}} - 
		\frac{2v^{a}_{ml}v^b_{nm}v^c_{ln}}{\epsilon^2_{lm}\epsilon^2_{nm}} \\
		&+\frac{4v^{a}_{nl}v^b_{mn}v^c_{lm}}{\epsilon^3_{nl}\epsilon_{nm}} + \frac{4v^{a}_{nl}v^b_{lm}v^c_{mn}}{\epsilon^3_{nl}\epsilon_{nm}} + 
		\frac{4v^{a}_{ln}v^b_{ml}v^c_{nm}}{\epsilon^3_{nl}\epsilon_{nm}} + 
		\frac{4v^{a}_{ln}v^b_{nm}v^c_{ml}}{\epsilon^3_{nl}\epsilon_{nm}}  \\
		&- \frac{4v^{a}_{lm}v^b_{mn}v^c_{nl}}{\epsilon^3_{lm}\epsilon_{ln}} - \frac{4v^{a}_{lm}v^b_{nl}v^c_{mn}}{\epsilon^3_{lm}\epsilon_{ln}} - 
		\frac{4v^{a}_{ml}v^b_{ln}v^c_{nm}}{\epsilon^3_{lm}\epsilon_{ln}} - 
		\frac{4v^{a}_{ml}v^b_{nm}v^c_{ln}}{\epsilon^3_{lm}\epsilon_{ln}}  \\ &-2\frac{v^b_{nl}v^c_{lm}v^a_{mn}}{\epsilon_{nl}\epsilon_{nm}^3} - 2\frac{v^c_{nl}v^b_{lm}v^a_{mn}}{\epsilon_{nl}\epsilon_{nm}^3} -
		2\frac{v^c_{ml}v^b_{ln}v^a_{nm}}{\epsilon_{nl}\epsilon_{nm}^3} - 
		2\frac{v^b_{ml}v^c_{ln}v^a_{nm}}{\epsilon_{nl}\epsilon_{nm}^3}  \\
		&+ 2\frac{v^c_{nl}v^b_{lm}v^a_{mn}}{\epsilon_{lm}\epsilon_{nm}^3} + 2\frac{v^b_{nl}v^c_{lm}v^a_{mn}}{\epsilon_{lm}\epsilon_{nm}^3} +
		2\frac{v^b_{ml}v^c_{ln}v^a_{nm}}{\epsilon_{lm}\epsilon_{nm}^3} + 
		2\frac{v^c_{ml}v^b_{ln}v^a_{nm}}{\epsilon_{lm}\epsilon_{nm}^3} \\
		& + \frac{2}{\epsilon^2_{nm}\epsilon_{ml}\epsilon_{ln}} \Big(v^a_{nl}v^{c}_{lm}v^b_{mn} + v^a_{nl}v^{b}_{lm}v^c_{mn} + v^c_{ml}v^{a}_{ln}v^b_{nm} + v^b_{ml}v^{a}_{ln}v^c_{nm}\Big) \\
		& - \frac{2}{\epsilon^2_{nm}\epsilon_{ml}\epsilon_{ln}} \Big(v^a_{ml}v^{c}_{ln}v^b_{nm} + v^c_{nl}v^{a}_{lm}v^b_{mn} + v^a_{ml}v^{b}_{ln}v^c_{nm} + v^b_{nl}v^{a}_{lm}v^c_{mn}\Big)
		\Big]\\ 
	\end{aligned}
\end{equation}
Next, we reclassify these multiband contributions according to their matrix element products, appropriately interchanging the dummy indices $m \leftrightarrow l$ where necessary. Upon summing the coefficients of these matrix products, we find that the total multiband contribution vanishes identically.
\begin{equation}
	\begin{aligned}
		\mathcal{M}
		& = \frac{1}{2} {\textstyle \sum_{nml}'} f_n\Big[ (2v^a_{lm}v^b_{mn}v^c_{nl} + 2v^a_{lm}v^c_{mn}v^b_{nl}) \Big(\frac{1}{\epsilon_{lm}^2 \epsilon_{nm}^2} + \frac{1}{\epsilon_{lm}^2 \epsilon_{nl}^2} + \frac{2}{\epsilon_{lm}^3 \epsilon_{ln}} - \frac{2}{\epsilon_{ml}^3 \epsilon_{nm}}  + \frac{1}{\epsilon_{nm}^2 \epsilon_{ml} \epsilon_{ln}} + \frac{1}{\epsilon_{nl}^2 \epsilon_{lm} \epsilon_{mn}}\Big) \\
		& \quad + (v^a_{ln}v^b_{nm}v^c_{ml} + v^{a}_{ln}v^c_{nm}v^b_{ml} + v^a_{nl}v^b_{lm}v^c_{mn} + v^{a}_{nl}v^c_{lm}v^b_{mn})\Big(-\frac{2}{\epsilon_{nl}^2\epsilon_{nm}^2} - \frac{4}{\epsilon_{nl}^3 \epsilon_{nm}} + \frac{2}{\epsilon_{nl}^3\epsilon_{nm}} - \frac{2}{\epsilon_{nl}^3\epsilon_{ml}} - \frac{2}{\epsilon_{nm}^2\epsilon_{ml}\epsilon_{ln}}\Big)
		\Big] \\
		& = 0
	\end{aligned}
\end{equation}

The vanishing of $\mathcal{M}$ signifies that multiband effects do not introduce any contributions to the final response. This is consistent with the case of the rectified current. Consequently, $T^{\text{SHG}}_0$ reduces to a combination of two-band terms:
\begin{equation}
	\begin{aligned}
		&T^{\text{SHG}}_{0} = T^{\text{SHG}}_{0, 1} + T^{\text{SHG; two-band}}_{0, 2+3} \\
		&=- \frac{1}{2} {\textstyle \sum_{nm}'} \frac{f_n}{\epsilon^3_{nm}}\Big[\partial_a(\ln A^{c}_{nm}) v^c_{nm} v^{b}_{mn} + \partial_a(\ln A^{b}_{nm}) v^b_{nm} v^{c}_{mn} + \partial_a(\ln A^{c}_{mn}) v^c_{mn} v^{b}_{nm} + \partial_a(\ln A^{b}_{mn}) v^b_{mn} v^{c}_{nm} \\
		&+ 2\partial_c(\ln A^{b}_{nm}) v^b_{nm} v^{a}_{mn} + 2\partial_c(\ln A^{b}_{mn}) v^b_{mn} v^{a}_{nm} + 2\partial_b(\ln A^{c}_{nm}) v^c_{nm} v^{a}_{mn} + 2\partial_b(\ln A^{c}_{mn}) v^c_{mn} v^{a}_{nm}\\
		&+2\partial_c(\ln A^{a}_{mn}) v^a_{mn} v^{b}_{nm} + 2\partial_c(\ln A^{a}_{nm}) v^a_{nm} v^{b}_{mn} + 2\partial_b(\ln A^{a}_{mn}) v^a_{mn} v^{c}_{nm} + 2\partial_b(\ln A^{a}_{nm}) v^a_{nm} v^{c}_{mn} \\
		&-2\partial_a(\ln A^{c}_{nm}) v^c_{nm} v^{b}_{mn} - 2\partial_a(\ln A^{b}_{nm}) v^b_{nm} v^{c}_{mn} - 2\partial_a(\ln A^{c}_{mn}) v^c_{mn} v^{b}_{nm} - 2\partial_a(\ln A^{b}_{mn}) v^b_{mn} v^{c}_{nm} \\
		& + 2\frac{\Delta_{nm}^a (v^b_{mn}v^c_{nm} + v^c_{mn}v^b_{nm})}{\epsilon_{nm}} + 5\frac{\Delta_{nm}^c (v^a_{mn}v^b_{nm} + v^b_{mn}v^a_{nm})}{\epsilon_{nm}} + 5\frac{\Delta_{nm}^b (v^a_{mn}v^c_{nm} + v^c_{mn}v^a_{nm})}{\epsilon_{nm}} \\
		&-\frac{\Delta_{nm}^a (v^b_{mn}v^c_{nm} + v^c_{mn}v^b_{nm})}{\epsilon_{nm}} - 7\frac{\Delta_{nm}^c (v^a_{mn}v^b_{nm} + v^b_{mn}v^a_{nm})}{\epsilon_{nm}} - 7\frac{\Delta_{nm}^b (v^a_{mn}v^c_{nm} + v^c_{mn}v^a_{nm})}{\epsilon_{nm}} \Big]\\
	\end{aligned}
\end{equation}
By invoking the Feynman-Hellmann identity, this expression can be further transformed into:
\begin{equation}
	\begin{aligned}
		&T^{\text{SHG}}_{0} 
		= \frac{1}{2} {\textstyle \sum_{nm}'} f_n \Big[\frac{\partial_{a}(A^{c}_{nm}A^{b}_{mn} + A^{b}_{nm}A^{c}_{mn})}{\epsilon_{nm}} -2 \frac{\partial_{b}(A^{a}_{nm}A^{c}_{mn} + A^{c}_{nm}A^{a}_{mn})}{\epsilon_{nm}}  -2 \frac{\partial_{c}(A^{a}_{nm}A^{b}_{mn} + A^{b}_{nm}A^{a}_{mn})}{\epsilon_{nm}} \\
		+& (A^{c}_{nm}A^{b}_{mn} + A^{b}_{nm}A^{c}_{mn}) \partial_a(\frac{1}{\epsilon_{nm}}) - 2(A^{a}_{nm}A^{c}_{mn} + A^{c}_{nm}A^{a}_{mn})\partial_b(\frac{1}{\epsilon_{nm}}) - 2(A^{a}_{nm}A^{b}_{mn} + A^{b}_{nm}A^{a}_{mn})\partial_c(\frac{1}{\epsilon_{nm}})\Big] \\
	\end{aligned}
\end{equation}
Finally, we observe that these partial derivatives can be reorganized into the form of total derivatives of the band-renormalized quantum metric:
\begin{equation}
	\begin{aligned}
		T^{\text{SHG}}_{0}
		&= -{\textstyle \sum_{n}} f_n (2\partial_c\mathcal{G}^{ab}_n + 2\partial_b\mathcal{G}^{ac}_n - \partial_a\mathcal{G}^{bc}_n)\\
	\end{aligned}
\end{equation}

\subsection{Transformation and Final Expressions}
Upon evaluating the low-frequency SHG conductivity, several intermediate terms appear formally distinct from those of the rectified conductivity. However, once summarized and categorized using a similar scheme, low-frequency SHG and rectified conductivities are revealed to be  identical. Here we employ IBP to simplify the expressions.

\subsubsection*{A. Kinetic Contribution}
The total intrinsic kinetic contribution is $\sigma^{\text{SHG}; \text{kin}}_{abc} = T_4 + T_3$. For $T_3$, we utilize the relation $4\partial_b\partial_c \epsilon_n \partial_a \epsilon_n = -2\partial_a(\partial_b \epsilon_n \partial_c \epsilon_n) + 2 \partial_b(\partial_a \epsilon_n \partial_c \epsilon_n) + 2\partial_c(\partial_a \epsilon_n \partial_b \epsilon_n) $ and perform IBP:
\begin{equation}
	\begin{aligned}
		T^{\text{SHG}}_3 &= -\frac{1}{2 \times 4!} {\textstyle \sum_{n}} f^{(3)}_n \Big[\partial_a\partial_c \epsilon_n \partial_b \epsilon_n + \partial_a\partial_b \epsilon_n \partial_c \epsilon_n + 4\partial_b\partial_c \epsilon_n \partial_a \epsilon_n\Big]
        \\
		&= -\frac{1}{2 \times 4!} {\textstyle \sum_{n}} f^{(3)}_n \Big[-\partial_a(\partial_b \epsilon_n \partial_c \epsilon_n) + 2 \partial_b(\partial_a \epsilon_n \partial_c \epsilon_n) + 2\partial_c(\partial_a \epsilon_n \partial_b \epsilon_n)\Big] \xrightarrow{\text{IBP}} \frac{3}{2 \times 4!} {\textstyle \sum_{n}} f^{(4)}_n v^a_n v^b_n v^c_n
	\end{aligned}
\end{equation}
Summing these contributions, we arrive at the final expression:
\begin{equation}
	\begin{aligned}
		\sigma^{\text{SHG}; \text{kin}}_{abc} &= -\frac{1}{4!} {\textstyle \sum_{n}} v^a_{n} v^{b}_{n}v^c_{n} f^{(4)}_n + \frac{3}{2 \times 4!} {\textstyle \sum_{n}} f^{(4)}_n v^a_n v^b_n v^c_n\\
		&= \frac{1}{48} {\textstyle \sum_{n}} v^a_{n} v^{b}_{n}v^c_{n} f^{(4)}_n,
        \quad
        f_n^{(4)} = 2 f_{0,n}^{(4)}
	\end{aligned}
\end{equation}

\subsubsection*{B. Geometric Contribution}
The rest of intrinsic contribution is geometric $\sigma^{\text{SHG}; \text{geo}}_{abc} = T_2 + T_1 + T_0$. 
By performing integration by parts (IBP) on $T^{\text{SHG}}_{0}$, we can recast it into a form analogous to $T_1$:
\begin{equation}
	\begin{aligned}
		T^{\text{SHG}}_{0}
		\xrightarrow{\text{IBP}}& {\textstyle \sum_{n}} f^{(1)}_n (2v^c_n\mathcal{G}^{ab}_n + 2v^b_n\mathcal{G}^{ac}_n - v^a_n\mathcal{G}^{bc}_n)\\
	\end{aligned}
\end{equation}
Likewise, performing IBP on $T_2$ yields:
\begin{equation}
	T^{\text{SHG}}_{2}
	\xrightarrow{\text{IBP}} -\frac{1}{2}{\textstyle \sum_{n}} f^{(1)}_n \partial_{a}g^{bc}_n
\end{equation}
Summing these three components, we obtain the total geometric contribution:
\begin{equation}
	\begin{aligned}
		\sigma^{\text{SHG}; \text{geo}}_{abc} &= -\frac{1}{2}{\textstyle \sum_{n}} f^{(1)}_n \partial_{a}g^{bc}_n -{\textstyle \sum_{n}} f^{(1)}_n \Big[v^b_{n} \mathcal{G}^{ac}_{n}  + v^c_{n} \mathcal{G}^{ab}_{n} + v^a_{n} \mathcal{G}^{bc}_{n} \Big] + {\textstyle \sum_{n}} f^{(1)}_n (2v^c_n\mathcal{G}^{ab}_n + 2v^b_n\mathcal{G}^{ac}_n - v^a_n\mathcal{G}^{bc}_n)\\
		&= {\textstyle \sum_{n}} f^{(1)}_n \Big[v^b_{n} \mathcal{G}^{ac}_{n}  + v^c_{n} \mathcal{G}^{ab}_{n} - 2v^a_{n} \mathcal{G}^{bc}_{n}  - \frac{\partial_a g^{bc}_n}{2}\Big],
        \quad
        f_n^{(1)} = 2 f_{0,n}^{(1)}
	\end{aligned}
\end{equation}

\section{Non-equivalence of the Fermionic Bath Model to the RTA/IFR Framework}

Although our results of the $\Gamma^0$ nonlinear conductivity for the wide-band fermionic bath share certain characteristics with the RTA/IFR framework, the discrepancies in coefficients and the emergence of extra contributions demonstrate that the two approaches are inherently non-equivalent. To substantiate this, we show that reducing our exactly solvable model to the phenomenological RTA/IFR form requires exceptionally crude  approximations. 

First, we derive the exact equation of motion for the reduced density matrix. Following the approach by Matsyshyn et al. \cite{matsyshyn2023fermi}, when considering a fermionic bath, the reduced density matrix can be expressed in the following form:
\begin{equation}
    \rho_S(t) = \sum_{n, j} f_0(\varepsilon_{j}) \ket{\psi^{j}_n (t)}\bra{\psi^{j}_n (t)}
\end{equation}
where $\ket{\psi^{j}_n (t)}$ is the component within the system Hilbert space evolving from the initial state of the fermionic bath $\ket{\phi_{n, j}}$. The exact equation of motion for $\ket{\psi^{j}_n (t)}$ is given by:
\begin{equation}
    i\partial_t \ket{\psi^{j}_n (t)} = [H_S(t) - i\Gamma]\ket{\psi^{j}_n (t)} + \lambda \exp [-i\varepsilon_j (t-t_0)] \ket{\chi_n} .
\end{equation}
The above equation is a simple non-Hermitian open-system Schrödinger equation, in which the system Hamiltonian is dressed by a constant imaginary part ``$-i \Gamma$'' which captures the decay into the bath. However, we see that the bath is not only to induce decay, but also produces the source term $\lambda \exp [-i\varepsilon_j (t-t_0)] \ket{\chi_n}$ that makes the equation inhomogeneous. The balance of decay and source term is what allows the existence of nontrivial non-equilibrium steady state.

From this, we derive the equation of motion for the reduced density matrix:
\begin{equation}
    \begin{aligned}
        i\partial_t \rho_S(t) &= i\sum_{n, j} f_0(\varepsilon_{j})\Big[\Big(\partial_t \ket{\psi^{j}_n (t)}\Big)\bra{\psi^{j}_n (t)} + \ket{\psi^{j}_n (t)}\Big(\partial_t \bra{\psi^{j}_n (t)}\Big)\Big] \\
        &= \sum_{n, j} f_0(\varepsilon_{j}) \Big[(H_S(t) - i\Gamma)\ket{\psi^{j}_n (t)}\bra{\psi^{j}_n (t)} + \lambda \exp [-i\varepsilon_j (t-t_0)] \ket{\chi_n}\bra{\psi^{j}_n (t)} \Big] \\
        &-\sum_{n, j} f_0(\varepsilon_{j}) \Big[\ket{\psi^{j}_n (t)}\bra{\psi^{j}_n (t)}(H_S(t) + i\Gamma) + \lambda \exp [i\varepsilon_j (t-t_0)] \ket{\psi^{j}_n (t)}\bra{\chi_n} \Big] \\
        &= [H_S(t), \rho_S(t)] - 2i\Gamma \rho_S(t) + \lambda \sum_{n, j}  f_0(\varepsilon_{j}) \Big[\exp [-i\varepsilon_j (t-t_0)] \ket{\chi_n}\bra{\psi^{j}_n (t)} - \text{h.c.}\Big] \\
        \equiv & [H_S(t), \rho_S(t)] - 2i\Gamma \rho_S(t) + {\cal S}[\psi^{j}_n (t)]
    \end{aligned}
\end{equation}
Again we see a source term $\mathcal{S}[\psi^{j}_n (t)]$ that makes the equation inhomogeneous. We examine the source term ${\cal S}[\psi^{j}_n (t)]$. To force this exact dynamic equation to reduce to the standard RTA/IFR form, one needs to apply two approximations:

First, one needs to apply a {\it static Hamiltonian approximation} to the source term. Specifically, one must artificially freeze the system Hamiltonian within the equations of motion, enforcing $H_S(t) \to H_S(0)$. This entirely discards the dynamic dressing of the states by the driving field. Under this assumption, a direct calculation yields:
\begin{equation}
    \ket{\psi^{j}_n (t)} = - \frac{\lambda}{\epsilon_n - i\Gamma - \varepsilon_j} e^{-i\varepsilon_j (t-t_0)} \ket{\chi_n}
\end{equation}
\begin{equation}
    \begin{aligned}
        {\cal S}[\psi^{j}_n (t)] & =  \sum_{n, j} f_0(\varepsilon_j) \frac{2i\lambda^2 \Gamma}{(\epsilon_n - \varepsilon_j)^2 + \Gamma^2} \ket{\chi_n}\bra{\chi_n} = \sum_{n}\int^{\infty}_{-\infty} d\omega_b f_0(\omega_b) \frac{2i\lambda^2 \Gamma \sum_{j} \delta(\varepsilon_j - \omega_b)}{(\epsilon_n - \omega_b)^2 + \Gamma^2} \ket{\chi_n}\bra{\chi_n} \\
        & = 2i\Gamma \sum_{n} \int^{\infty}_{-\infty} \frac{d\omega_b}{\pi}  f_0(\omega_b) \frac{\Gamma}{(\epsilon_n - \omega_b)^2 + \Gamma^2} \ket{\chi_n}\bra{\chi_n}
    \end{aligned}
\end{equation}
Here, we have employed the assumption of a featureless fermionic bath, characterized by the spectral density $\nu_B = \sum_{j} 2\pi \delta(\varepsilon_j - \omega_b)$, along with the definition of the relaxation rate $\Gamma = \lambda^2 \nu_B / 2$.

Second, one needs to further perform a {\it relaxation rate truncation}. By expanding the dissipation parameter around $\Gamma = 0$ and strictly retaining only the first-order term, one effectively replaces the exact Lorentzian spectral broadening with a Dirac delta function:
\begin{equation}
    \begin{aligned}
        {\cal S}[\psi^{j}_n (t)] & = 2i\Gamma \sum_{n} \int^{\infty}_{-\infty} \frac{d\omega_b}{\pi} f_0(\omega_b) \frac{\Gamma}{(\epsilon_n - \omega_b)^2 + \Gamma^2} \ket{\chi_n}\bra{\chi_n} \\
        & =  2i\Gamma \sum_{n} f_0(\epsilon_n) \ket{\chi_n}\bra{\chi_n} + O(\Gamma^2) = 2i\Gamma \rho_0 + O(\Gamma^2)
    \end{aligned}
\end{equation}
We observe that ${\cal S}[\psi^{j}_n (t)]$ reduces to $2i\Gamma \rho_0 + O(\Gamma^2) $, which is proportional to the density matrix of the fermionic system at equilibrium when all higher-order $\Gamma$ contributions are ignored. These higher-order contributions can be systematically extracted by expressing the integral as a convolution and expanding its Fourier-space representation via the Taylor series of $e^{-\Gamma|k|}$:
\begin{align}
\int \frac{d \omega_b}{\pi} f_0\left(\omega_b\right) \frac{\Gamma}{\left(\epsilon_n-\omega_b\right)^2+\Gamma^2}=\frac{1}{2 \pi} \int d k \tilde{f}_0(k) e^{i k \epsilon_n} e^{-\Gamma|k|} 
\end{align}

Applying these two approximations reduces the source term to ${\cal S}[\psi^{j}_n (t)] \simeq 2i\Gamma \rho_0$. Defining the relaxation time $\tau = (2\Gamma)^{-1}$, the equation of motion for the reduced density matrix simplifies to the phenomenological RTA form:
\begin{equation}
    \begin{aligned}
        \partial_t \rho_S(t) &= -i[H_S(t), \rho_S(t)] - \frac{\rho_S(t) - \rho_0}{\tau} \\
    \end{aligned}
\end{equation}
This reproduces the quantum Liouville equation used by RTA/IFR framework \cite{das2023intrinsic}. However, as previously discussed, these approximations are crude for nonlinear transport. The first approximation artificially deletes all Floquet information at finite AC frequencies, while the second truncation ignores all higher-order $O(\Gamma^2)$ contributions generated by the drive and the bath. Consequently, our exact Markovian fermionic bath model cannot be reduced to the RTA/IFR framework which is often employed to describe impurity scatterings without discarding the specific mechanisms that actively shape the observable intrinsic conductivity.

\clearpage

\end{document}